\documentclass[graybox]{svmult}

\usepackage{mathptmx}       
\usepackage{helvet}         
\usepackage{courier}        
\usepackage{type1cm}        
\usepackage{makeidx}         
\usepackage{graphicx}        
\usepackage{multicol}        
\usepackage[bottom]{footmisc}

\newcommand{\tn}{\tabularnewline}

\usepackage{balance}
\usepackage{algorithm,algorithmic}
\usepackage{multirow}
\usepackage{appendix}
\usepackage{array}
\usepackage{ctable}
\usepackage[size=small,format=plain,labelfont=up,textfont=up]{caption}
\DeclareCaptionType{copyrightbox}
\usepackage{caption}
\usepackage{subcaption}
\usepackage{filecontents}
\usepackage{amsmath,amssymb}
\usepackage{epstopdf}
\usepackage{hyperref}
\usepackage{mathtools}
\usepackage{wrapfig}

\usepackage{hyperref}

\makeindex

\begin{document}

\title*{Location Privacy in Spatial Crowdsourcing}
\author{Hien To and Cyrus Shahabi}

\institute{Hien To \at University of Southern California, Los Angeles, CA 90089 \email{hto@usc.edu}
\and Cyrus Shahabi \at University of Southern California, Los Angeles, CA 90089 \email{shahabi@usc.edu}}
%
%
\maketitle

\abstract{
Spatial crowdsourcing (SC) is a new platform that engages individuals in collecting and analyzing environmental, social and other spatiotemporal information. With SC, requesters outsource their spatiotemporal tasks (tasks associated with location and time) to a set of workers, who will perform the tasks by physically traveling to the tasks' locations. However, current solutions require the workers, who in many cases are simply volunteering for a cause, to disclose their locations to untrustworthy entities. Revealing an individual's location data to other entities may prevent people from contributing to SC applications, thus rendering location privacy a critical obstacle to the growth of SC applications.
\\
This chapter first identifies privacy threats toward both workers and requesters during the two main phases of spatial crowdsourcing, tasking and reporting. \emph{Tasking} is the process of identifying which tasks should be assigned to which workers. This process is handled by a spatial crowdsourcing server (SC-server). The latter phase is \emph{reporting}, in which workers travel to the tasks' locations, complete the tasks and upload their reports to the SC-server.
The challenge is to enable effective and efficient tasking as well as reporting in SC without disclosing the actual locations of workers (at least until they agree to perform a task) and the tasks themselves (at least to workers who are not assigned to those tasks). 
\\
This chapter aims to provide an overview of the state-of-the-art in protecting users' location privacy in spatial crowdsourcing. We provide a comparative study of a diverse set of solutions in terms of task publishing modes (push vs. pull), problem focuses (tasking and reporting), threats (server, requester and worker), and underlying technical approaches (from pseudonymity, cloaking, and perturbation to exchange-based and encryption-based techniques). The strengths and drawbacks of the techniques are highlighted, leading to a discussion of open problems and future work.
}

\section{Introduction}

The increase in computational and communication performance of mobile devices, coupled with the advances in sensor technology, leads to an exponential growth in data collection and sharing by smartphones. Exploiting this large volume of potential users\footnote{Throughout the chapter we use ``user" when refering to both worker and requester.} and their mobility, a new mechanism for efficient and scalable data collection has emerged, namely, \textit{spatial crowdsourcing} (SC)~\cite{Kazemi2012}.
SC has numerous applications in domains such as environmental sensing (iRain\footnote{irain.eng.uci.edu}~\cite{iRain}), smart cities (Waze\footnote{waze.com} and TaskRabbit\footnote{taskrabbit.com}), journalism and crisis response (MediaQ\footnote{mediaq.usc.edu}~\cite{kim2014mediaq}). To illustrate, consider a disaster-response scenario where the Red Cross (i.e., requester) is interested in collecting pictures and videos of disaster areas from various locations of a city. With SC, the requester issues a query to a crowdsourcing server (SC-server). The SC-server then distributes the query among the available workers in the vicinity of the events. Once the workers document their events with their mobile phones, the results are sent back to the requester.
Typically, requesters and workers register with the SC-server that acts as a broker between parties, and often also plays a role in how tasks are assigned to workers (i.e., scheduling according to some performance criteria). 
We refer to this phase as \emph{tasking} (aka task assignment). After tasking, workers travel to the locations of the tasks, perform them and report the results to the SC-server. This phase is referred to as \emph{reporting}.


Both tasking and reporting phases often require workers and requesters to reveal locations of workers and tasks to potentially untrustworthy entities (SC-server, other workers and other requesters). Effective tasking is an important phase in spatial crowdsourcing so that tasks are completed in a timely fashion, and workers do not incur significant travel cost~\cite{Kazemi2012,Kazemi2013,Deng2013,To2016}. Hence, matching (or worker selection) must take into account the locations of workers and tasks, revealing private locations to the SC-server. Similarly, reporting spatial tasks would enable the SC-server to infer the workers' locations since they must have visited the locations of the tasks. However, disclosing individual locations has serious privacy implications. Leaked locations are often collected and shared without user consent~\cite{angwin2011,mcmillan2014}, leading to a breach of sensitive information such as an individual's health (e.g., presence in a cancer treatment center), alternative lifestyles, political and religious preferences (e.g., presence in a church). Knowing user locations, an adversary can stage a broad spectrum of attacks such as physical surveillance and stalking, and identity theft~\cite{scheck2010}.
Particularly, in~\cite{Wang2016}, the authors show that hackers can stalk users in Waze---a popular SC application---by generating fake events such as accidents.
Consequently, mobile users may not agree to engage in spatial crowdsourcing if their privacy is violated;
thus, ensuring location privacy is key to the success of SC.
One may argue that simply removing identities of workers and tasks by using fake identities (i.e., pseudonymity) can achieve privacy. However, hiding users' identities without hiding their locations is inadequate.
The user's \emph{location trace} can be easily associated with a certain residence or office, which reveals the user's identity.
Hence, hiding a worker's location is much more challenging than hiding his/her identity.

Location privacy has been studied before in the context of location-based services. Several solutions~\cite{gruteser03,mca06,kalnis2007preventing} have been proposed to protect location-based queries, i.e., given a user's location, find points of interest in the proximity without disclosing the actual coordinates. However, in SC, the worker location is no longer part of the query, but rather the result of a spatial query around the task location. 
In addition, while some studies consider queries on private locations in the context of outsourced databases~\cite{ylx13,choi2014secure}, it is assumed that the data owner entity and the querying entity trust each other, with protection being offered only against intermediate service provider entities. This scenario does not apply in SC, as there is no inherent trust relationship between requesters and workers.
In the most restrictive privacy settings, all SC parties could be hostile to one another.

The first step of the tasking phase is task publication. There are two modes of task publication in SC: push (e.g., iRain) vs. pull (e.g., TaskRabbit). With the pull mode, the SC-server publishes the spatial tasks and online workers can choose any spatial task in their vicinity without the need to coordinate with the server. With the push mode, online workers send their locations to the SC-server, which then assigns to every worker his nearby tasks (posted by requesters). Each mode shares similar challenges and has its own unique challenge. The common challenges are that a worker should know a task location only if he plans to perform the task; likewise, only requesters who have tasks performed by the worker should know his location. Furthermore, the unique challenge with the push mode is that the SC-server must match workers to tasks without compromising their privacy. This requires strategies to ensure effective task assignment without revealing locations of tasks and workers. On the other hand, the unique challenge with the pull mode is to enable every worker to request tasks, perform them and subsequently post the results to the SC-server without revealing his location and identity.
Finally, providing privacy protection simultaneously both tasking and reporting phases introduces another set of challenges to both push and pull modes.

\begin{wraptable}{r}{0.45\textwidth}
\caption{Attacks on SC users.}
\begin{center}
\begin{tabular}{ | c | c | c | } 
\hline
					& Tasking 				& Reporting \\
\hline
Push 	& \cite{Kazemi2011a} 	& \cite{Shin2011}  \\  
\hline
Pull 	&  	[Sec.~\ref{sec:tasktabbit}]						& \cite{Wang2016,Shin2011}, [Sec.~\ref{sec:tasktabbit}] \\ 
\hline
\end{tabular}
\label{tab:threats}
\end{center}
\vspace{-15pt}
\end{wraptable}


Among the two modes of task publishing, privacy protection in the push mode is more challenging because tasking in the push mode is more complex than that of the pull mode. Countermeasure studies in the pull mode have been the main focus in the past decade with an emphasis on a special class of SC, named \emph{participatory sensing} (PS). PS usually assumes the pull mode of task publication (workers choose tasks); therefore, the main privacy threats to workers occur during reporting. Meanwhile, the most recent studies in SC have focused on the push mode (SC-server assigns tasks to workers); for this reason, main privacy breaches occur during tasking~\cite{Kazemi2011a}. Consequently, the existing studies in SC can be classified into two groups: 1) preserving privacy during \emph{reporting in the pull mode }~\cite{Shin2011,Boutsis2013,Zhang2016}, and 2) preserving privacy when \emph{tasking in the push mode }~\cite{Kazemi2011a,Vu2012,To2014,Pournajaf2014,Gong2015,Zhang2015,To2016b,Hu2015,Shen2016}.


In this chapter we study the privacy threats
to workers and requesters\footnote{Throughout this chaper we use ``requester" when referring to privacy threats toward a person and use ``task" when referring to privacy threats which would reveal the requester's location} in SC, during both tasking and reporting phases with either push or pull mode. Throughout this chapter we also identify three major drawbacks of the existing studies. First, they solely focus on protecting privacy during either phase of tasking or reporting, but not both. Second, none of these studies ensure privacy for both workers and requesters. To elaborate, we perform a set of simple attacks on TaskRabbit to demonstrate that locations of workers and requesters can be learned during both tasking and reporting phases. Third, despite the fact that most studies focus on either reporting in the pull mode or tasking in the push mode, privacy threats to SC users may also occur in other scenarios. Table~\ref{tab:threats} shows that there have been known attacks under the tasking and reporting phases with either the push or pull mode of task publishing. We demonstrate such threats in Section~\ref{sec:tasktabbit} via another set of attacks on TaskRabbit. These observations open some new research questions such as: how do we protect location privacy of both workers and tasks, simultaneously, during both the tasking and reporting phases of SC, and what are the promising privacy techniques to be used?

There have been recent surveys in privacy-preserving participatory sensing~\cite{Christin2016a} and mobile crowdsourcing~\cite{Pournajaf2015a}. Unlike these surveys, which provide an overview of a broad range of related problems, this chapter provides an in-depth study of the privacy challenges and the solutions proposed in the prior studies.
The remainder of this chapter is organized as follows. In Section~\ref{sec:1} we introduce spatial crowdsourcing and compare it with related concepts. Section~\ref{sec:2} illustrates potential privacy risks to both workers and requesters. Section~\ref{sec:3}  summarizes existing solutions addressing the privacy concerns in both the tasking and reporting phases of SC. Finally, we present our conclusions and future research directions in Section~\ref{sec:conclude}.

\section{Spatial Crowdsourcing}
\label{sec:1}

In this section we define spatial crowdsourcing and present two modes of task publishing, push vs. pull, with the push mode recently being dominant in the research community. Thereafter, we differentiate SC from the related topic of participatory sensing, which usually assumes the pull mode of task publication.

\paragraph{\textbf{Generic Framework}}

Spatial crowdsourcing (SC)~\cite{Kazemi2012} is a type of online crowdsourcing where performing tasks requires workers to physically be present at the locations of the tasks, termed \emph{spatial tasks}. A spatial task is a query to be answered at a particular location and must  be performed before a deadline. An example of a spatial task is taking a picture of a particular dish in a restaurant. This means that the workers need to physically travel to the location of the restaurant in order to take the picture.
A worker is a carrier of a mobile device who will perform spatial tasks for some incentives.

Spatial crowdsourcing (SC) has gained popularity in both the research community (e.g., \cite{Kazemi2012,To2014}) and industry (e.g., TaskRabbit, Gigwalk). A recent study \cite{to2015server} distinguishes SC from related fields, such as generic crowdsourcing, participatory sensing, volunteered geographic information, and online matching. Research efforts have focused on different aspects of SC, including task assignment, task scheduling, privacy, trust and incentive mechanism.

\paragraph{\textbf{Task Assignment: The Focus of Spatial Crowdsourcing}}

The main challenges of spatial crowdsourcing are due to the large-scale, ad hoc and dynamic nature of the workers and tasks.  To continuously match thousands of spatial crowdsourcing campaigns, where each campaign consists of many spatiotemporal tasks with millions of workers, an SC-Server must be able to run efficient task assignment (aka tasking). According to~\cite{Kazemi2012}, there are two types of tasking modes based on how workers are matched to tasks---server-assigned tasks (SAT) and worker-selected tasks (WST)--- which are also known as push and pull modes, respectively. Depending on the choice of a particular mode, the  focus of privacy protection is either at the tasking or the reporting stage of spatial crowdsourcing.

With the \emph{pull} mode, the SC-server publicly\footnote{Exact geographical coordinates of the tasks may not be published; instead, their cloaked locations or representative names are provided.} publishes the spatial tasks, and online workers autonomously choose tasks in their vicinity without coordinating with the SC-server. One advantage of the pull mode is that the workers do not need to reveal their locations to SC-server. However, one drawback of this mode is that the server does not have any control over the allocation of spatial tasks; this may result in some spatial tasks never be assigned, while others are assigned redundantly. Another drawback of the pull mode is that workers choose tasks based on their own objectives (e.g., choosing the $k$ closest spatial tasks to minimize their travel cost), which may not result in a globally optimal assignment. An example of the pull mode is TaskRabbit, where the workers browse for available spatial tasks and pick the ones in their neighborhood.

With the \emph{push} mode, requesters post tasks that include locations while online workers send their locations to the SC-server, which assigns tasks to nearby workers. The advantage of this mode is that unlike the pull mode, the SC-server has the big picture and can therefore assign to every worker his nearby tasks while maximizing the overall task assignment. However, the drawback is that locations of both tasks and workers should be sent to the server for effective assignment, which can pose privacy threats. An example of the push mode of SC is iRain~\cite{To2016}---a crowdsourcing framework that enables researchers to request rainfall information at specific locations and times where traditional means (e.g., satellite remote sensing and radar detection) fail to provide real-time, fine-grained data. Individual iRain users in the nearby locations can respond to those requests by reporting rainfall observations, such as heavy/medium/light/none.

Most SC studies assume the push mode and thus emphasize privacy protection during the tasking phase. With the pull mode, the main focus of privacy protection is shifted to the reporting phase, which has been well studied in the context of participatory sensing (e.g., ~\cite{Shin2011,Kazemi2011a,Vu2012,Boutsis2013,Zhang2016}). With participatory sensing, the goal is to exploit the ability of mobile users to collect and share data using their sensor-equipped phones for a given campaign. Most studies on participatory sensing focus on small campaigns with a limited number of workers; hence, they do not have issues of task assignment. However, with spatial crowdsourcing, the focus is on devising a scalable, generic and multipurpose crowdsourcing framework, similar to Amazon's Mechanical Turk, but spatial, where multiple campaigns can be handled simultaneously. Therefore, the main challenge with spatial crowdsourcing is to devise an efficient approach to assign tasks to workers given the large scale of an environment.

\section{Privacy Threats}
\label{sec:2}

There have been known attacks on SC applications; these include location-based attacks during tasking in the push mode~\cite{Kazemi2011a} and collusion attacks during reporting in the pull mode~\cite{Wang2016} (see Table~\ref{tab:threats}). Despite the fact that most studies have solely focused on one of the two major threats, privacy risks to SC users may occur in the other scenarios: reporting in the push mode and tasking in the pull mode.
In this section we present a threat model which characterizes the \emph{full spectrum of privacy threats to workers and requesters during both tasking and reporting phases with either push or pull mode}. Next, we illustrate the privacy risks by performing simple attacks on TaskRabbit.

\subsection{Threat Model}
\label{sec:threat_model}
As the privacy threats vary according to the modes of task publishing, we discuss possible threats associated with each mode.

\paragraph{\textbf{Privacy Threats with the Push Mode}}

\begin{figure*}[ht]
	\begin{minipage}[b]{0.49\linewidth}
		\centering
		\includegraphics[width=1\textwidth]{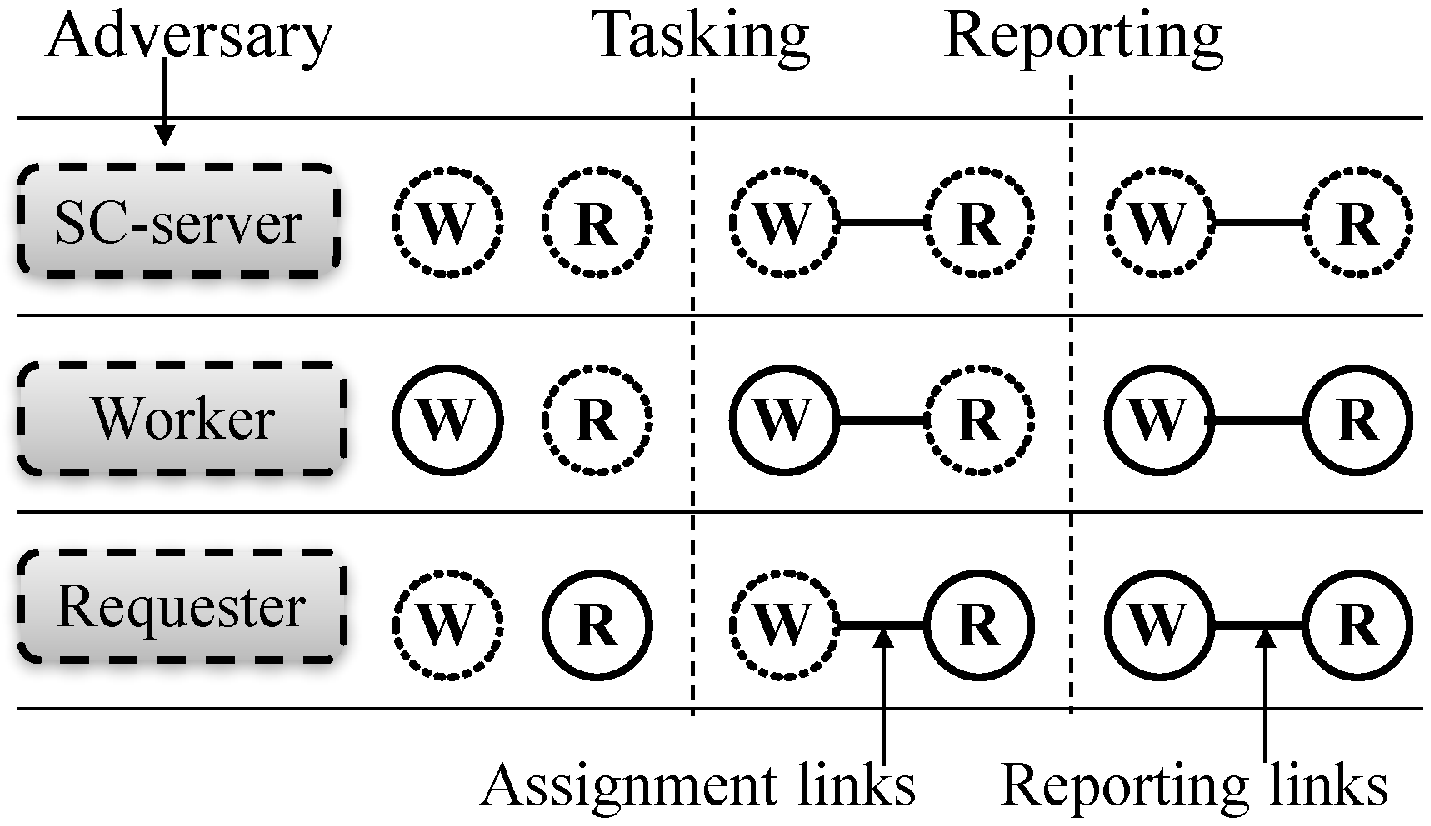}
		\subcaption{Push mode}
		\label{fig:threat_model_push}
	\end{minipage}
	\hspace{2pt}
	\begin{minipage}[b]{.49\linewidth}
		\centering
		\includegraphics[width=1\textwidth]{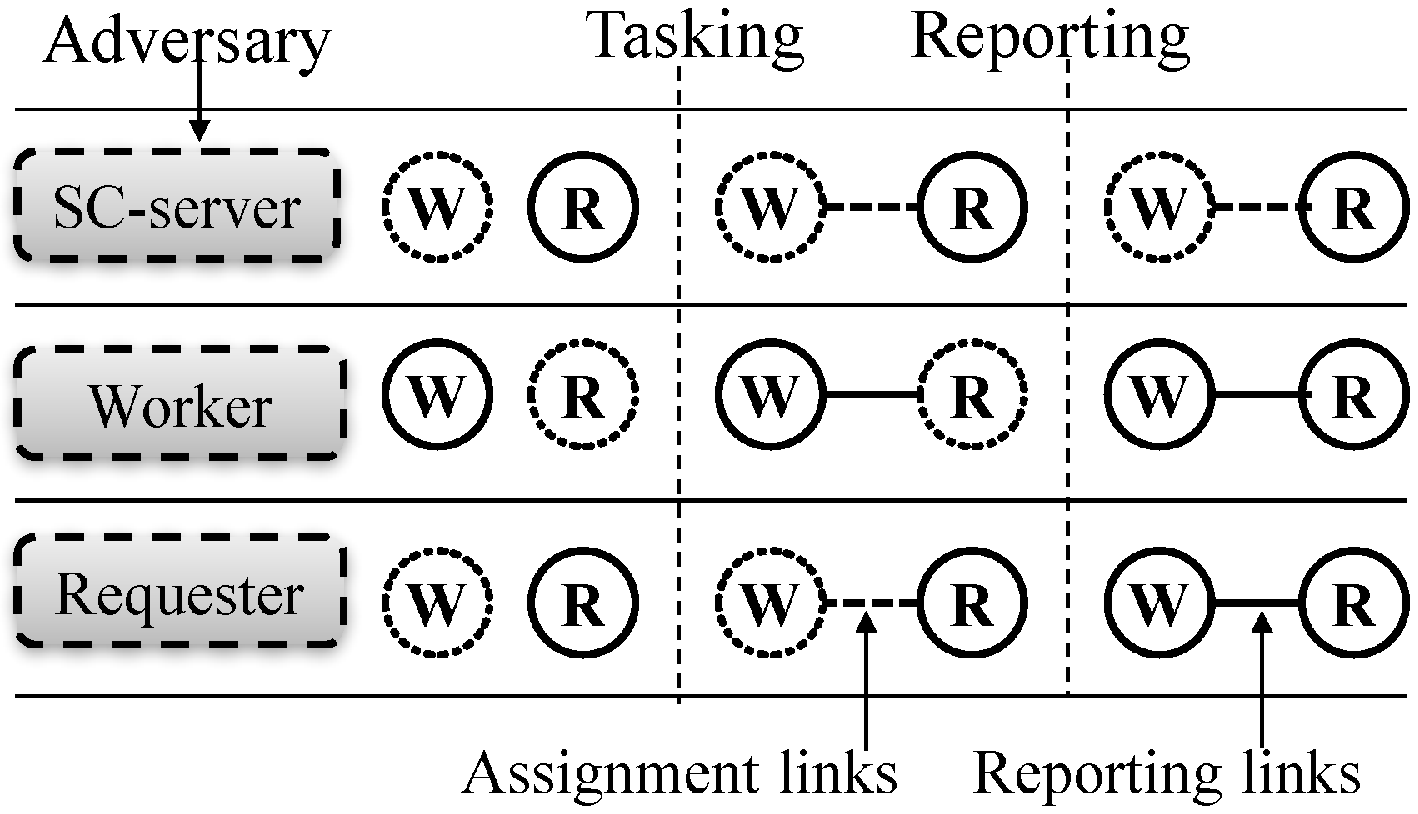}
		\subcaption{Pull mode}
		\label{fig:threat_model_pull}
	\end{minipage}
	\caption{Threat models in spatial crowdsourcing. W and R denote workers and requesters, respectively. The dotted circles surrounding them denote that they are protected from a malicious entity shown in the first column of the first row in a dashed shaded box.
		After the tasking and reporting phases, the links between W and R represent the established connections during each phase. We refer to these links as the \emph{assignment link} and \emph{reporting link}. The dashed links indicate connections that are oblivious to the corresponding malicious entity.}
	\label{fig:threat_model}
\end{figure*}

With the \emph{push} mode, SC-server takes as input locations of workers and tasks to perform effective task assignment; hence, there is a serious privacy threat from the SC-server which might become a single point of attack. Figure~\ref{fig:threat_model_push} depicts the threat model for the push mode of spatial crowdsourcing.
The first row means that locations of workers and tasks are protected from the SC-server at all the time. The role of the SC-server is to create the \emph{assignment links} between the workers and the requesters so that they can establish a direct communication channel among themselves. Each worker-requester pair cooperatively decides whether to accept the assignment from the SC-server. If yes, they send a \emph{consent} message to the SC-server, confirming that the worker will perform the requester's tasks. This agreement is illustrated by the first \emph{reporting link} in Figure~\ref{fig:threat_model_push}. We argue that to preserve location privacy during both tasking and reporting phases, task locations need to be protected from the SC-server. Otherwise, the completion of a task reveals that some workers must have visited the task's location. In restrictive privacy settings, workers and requesters can also be malicious to each other. Hence, to ensure minimum disclosure among them, in our threat model only workers who aim to perform the tasks should know the tasks' locations (see the second row in Figure~\ref{fig:threat_model_push}). Likewise, a requester should only know the workers' locations once her tasks are matched to and then performed by those workers (see the third row in Figure~\ref{fig:threat_model_push}). 

We emphasize that this threat model guarantees minimum disclosure of location information for both workers and tasks. The reason for this is twofold. First, the SC-server knows only the assignment links between workers and tasks. Due to such links, the assigned workers (or tasks) may infer that there exists nearby tasks (or workers). These disclosures are unavoidable in the push mode of SC. Second, the disclosure of workers' locations to their corresponding requester is inevitable at the reporting phase per definition of SC. It is worth mentioning that this threat model is restrictive; hence, weaker variants exist. For example, most existing studies in the push mode assume that workers are trusted~\cite{Kazemi2011a,Pournajaf2014,Hu2015} and task locations are public~\cite{Vu2012,To2014,Gong2015,Zhang2015,Shen2016}.

\paragraph{\textbf{Privacy Threats with the Pull Mode}}

With the \emph{pull} mode, despite the fact that workers do not need to send their locations to the SC-server, the locations can still be learned during both tasking and reporting phases. As long as a worker connects to the server to either \emph{request} some tasks or \emph{report} results, he may reveal to the server patterns of where and when the connections were made and what kind of tasks he wants to perform. Consequently, in~\cite{Shin2011}, the authors show that linking multiple requests or reports of the worker may allow an adversary to trace him since the worker's location information can be tracked through several stationary connection points (e.g., cell towers).
In addition, the worker's location trace can be inferred by both the SC-server and requesters since he must be in the vicinity of the tasks in order to perform them.
Figure~\ref{fig:threat_model_pull} depicts the proposed threat model for the pull mode. To preserve privacy and identity of the workers from the SC-server, both assignment links and reporting links should be secure during tasking and reporting phases, respectively. This is because if the connections are discovered by the SC-server, which already knows the locations of tasks, the server learns the locations of workers  since they must have visited the locations of the performed tasks. Hence, the workers must request tasks without revealing their identity to the SC-server; once the tasks are performed, the workers must also disassociate their connections with the performed tasks while uploading task content to the server.
Similar to the push mode, both workers and requesters themselves can be hostile to one another.
Thus, the privacy threats from workers and requesters (rows 2 and 3 in Figure~\ref{fig:threat_model_pull}) are similar to those in the push mode (rows 2 and 3 in Figure~\ref{fig:threat_model_push}), except the difference in the assignment links of the two second rows. The reason for this is that the requester is oblivious to the requests between the worker and the SC-server during tasking.


\subsection{Case Study of TaskRabbit}
\label{sec:tasktabbit}
We show that an adversary can perform harmful attacks on a typical SC application without much effort. TaskRabbit is a pull-based\footnote{We present the privacy threats to a pull-based SC system only; however, some of these privacy threats also occur in push-based SC such as iRain.} online and mobile marketplace that matches workers with requesters, allowing requesters to find immediate help with everyday tasks including, but not limited to, cleaning, moving, and delivery. In the following we discuss the aforementioned threats to TaskRabbit  users. Note that the following attacks on TaskRabbit.com were conducted in October 2014; the website has been updated since then.

\begin{figure*}[ht]
	\begin{minipage}[b]{.245\linewidth}
		\centering
		\includegraphics[width=1\textwidth]{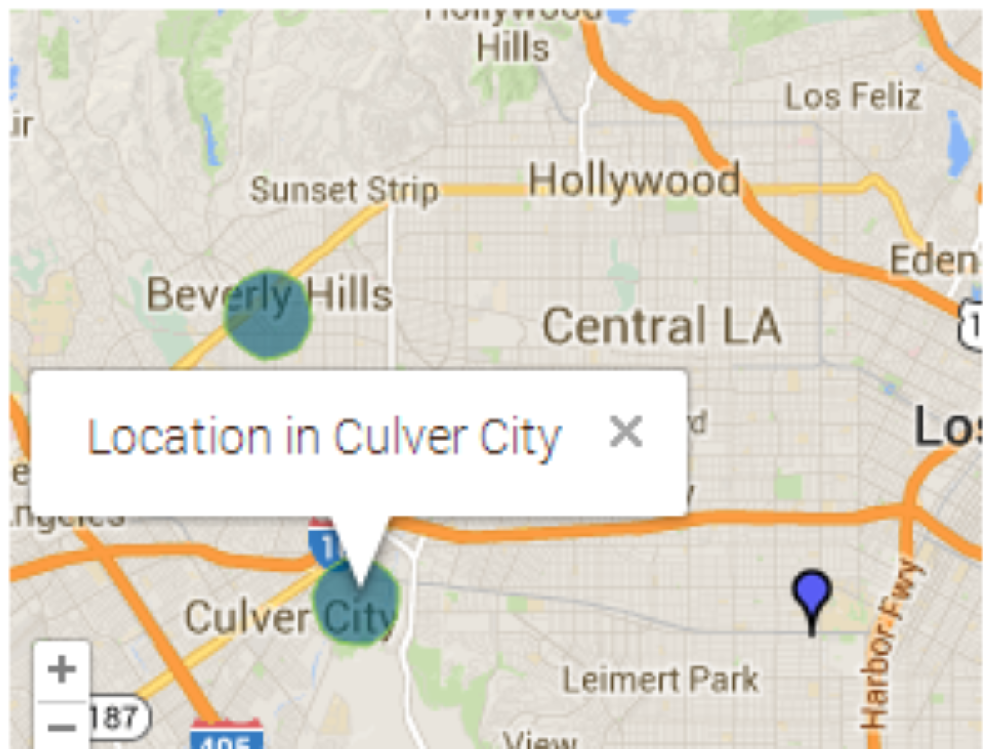}
		\subcaption{Task locations}
		\label{fig:task_location}
	\end{minipage}
		\begin{minipage}[b]{.245\linewidth}
		\centering
		\includegraphics[width=1\textwidth]{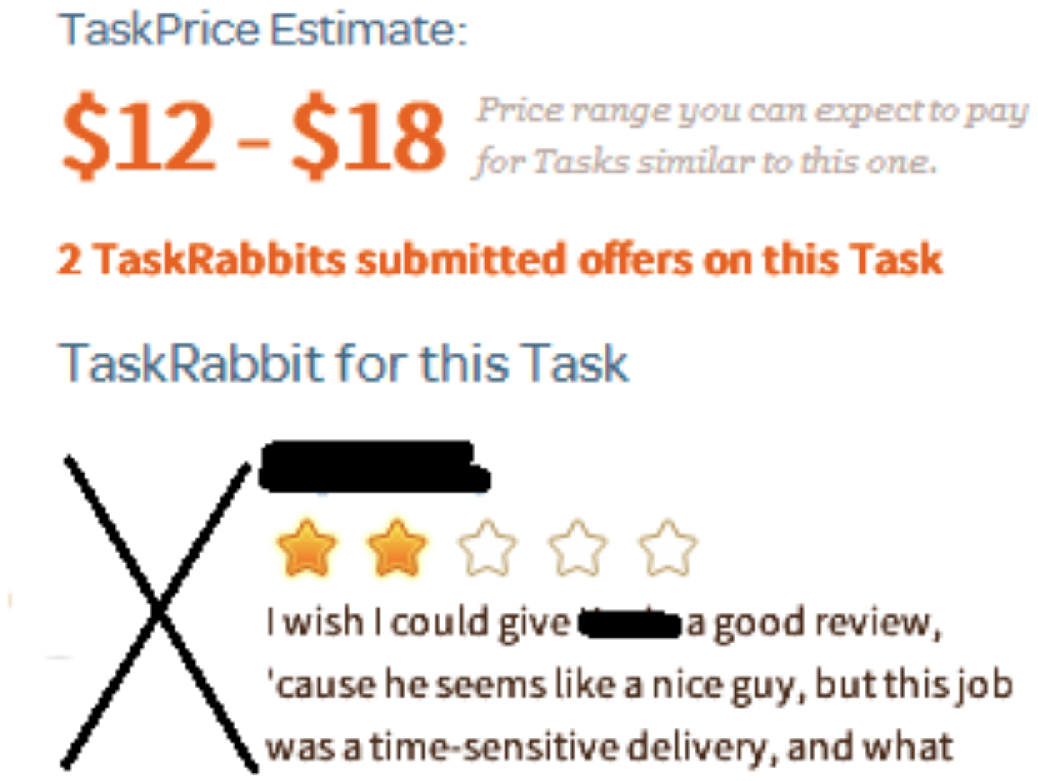}
		\subcaption{Task price}
		\label{fig:task_price}
	\end{minipage}
	\begin{minipage}[b]{.245\linewidth}
		\centering
		\includegraphics[width=1\textwidth]{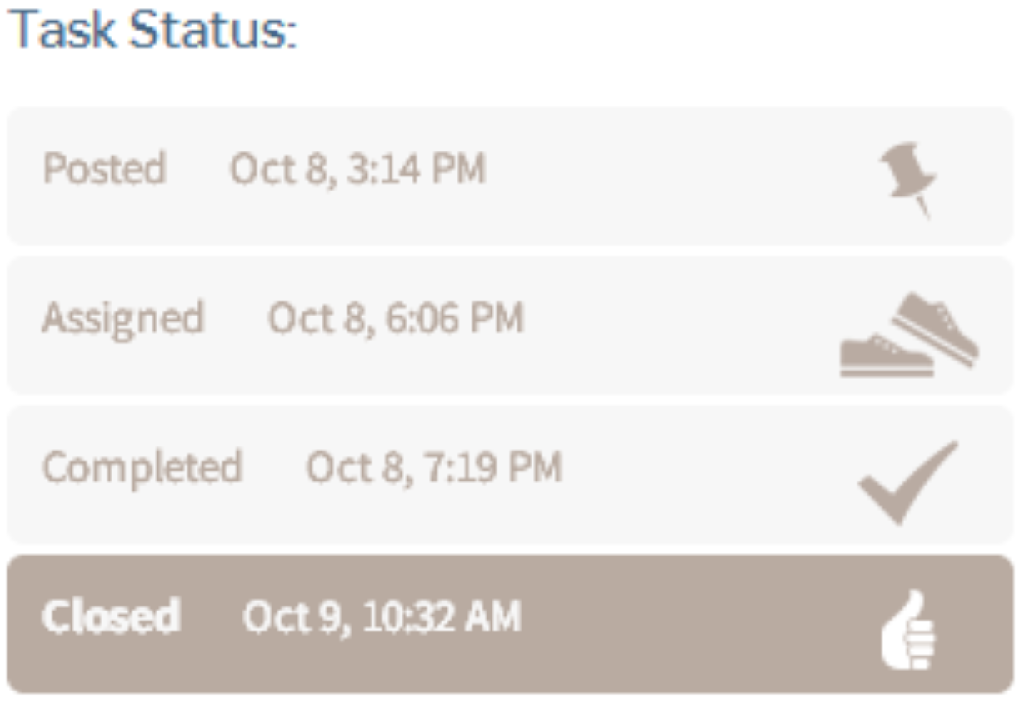}
		\subcaption{Task status}
		\label{fig:task_status}
	\end{minipage}
	\begin{minipage}[b]{.245\linewidth}
		\centering
		\includegraphics[width=1\textwidth]{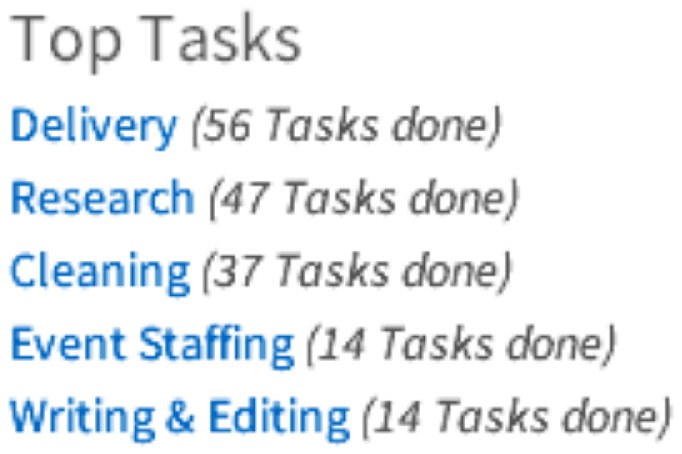}
		\subcaption{Performed tasks}
		\label{fig:top_tasks}
	\end{minipage}
	\caption{Screenshots of TaskRabbit web application from worker Bob.}
	\label{fig:task_info}
\end{figure*}

We first show the breach of task location during tasking. 
We signed up as a worker account and searched for delivery tasks in Los Angeles; 2381 spatial tasks were found. We obtained various information about a particular task by clicking on it, such as description, price, task status and cloaked locations. Although each location is cloaked in a circle with a radius of half a km\footnote{We obtained this information via JavaScript code.} (Figure~\ref{fig:task_location}) to protect task locations from workers, the actual drop-off and pick-up locations were mentioned in the task description, i.e., \emph{``Please pick up a box of mini-muffins from (S) promptly at 8 am on Tues, 9/4, and drive them straight to me at (D)."}
It is also worth noting that task requests often contain sensitive information, such as health status of the requesters. An example of a sensitive task is one with title \emph{``super easy task deliver a bag to the doorstep of a sick friend."}
Nonetheless, these privacy risks are due to the disclosure of task content, which is beyond the scope of this study.

We then show the leak of worker location during tasking and reporting. To gain a competitive advantage, a worker may wish to not disclose locations of his visits to other workers and requesters.
The task status (Figure~\ref{fig:task_status}) infers that the worker, referred to as Bob, was at the pick-up and drop-off locations of the task during the one-hour period between his assigned time and his completed time. The risk of precisely inferring Bob's locations is even higher for time-sensitive tasks such as delivery and help at home, which requires him to meet requesters in-person at a specific place and time. This inference attack shows that TaskRabbit does not guarantee privacy protection with respect to the threat model for the pull mode in Section~\ref{sec:threat_model}, which says that Bob's locations are private to the SC-server and only requesters who have their tasks performed by Bob should know his locations.
In addition, one can also see much more information about Bob, including his previously performed tasks (Figure~\ref{fig:top_tasks}) and all reviews from the requesters who hired him. These associations between Bob and his performed tasks indicate that the assignment links and reporting links are known to the SC-server, violating the threat model.

Among Bob's requesters, we randomly picked one named Alice.
We further show that her home location can be learned by tracking her task requests.
We searched for household tasks that Alice requested in the past; three of them are shown in Table~\ref{tab:requested_tasks}. These tasks were in the proximity of each other and likely situated at her home. Our hypothesis is that the tasks' locations were randomly cloaked such that the cloaking regions covered the actual location of the tasks.
The location must be in the overlapped area using triangulation. We validated our hypothesis by confirming that the location of another task, whose location was known, is within the overlapped region. This attack suggests that the more task requests are posted, the more accurately their locations can be learned. This simple attack is against the threat model, which states that the locations of Alice's tasks should only be revealed to the workers who performed her tasks.

\begin{table}
\begin{center}
            \caption{Three tasks requested by requester Alice. We replaced six digits after the decimal point of ``geo\_center" by 'x' to protect the privacy of the requester.}
    \begin{tabular}{ p{5cm}  |   p{6cm}}
    \hline
	Task description & Corresponding JavaScript \\ \hline
    Quick post-party dishwashing clean up needed & \small{``radius" : ``0.5", ``geo\_center" : \{``lat" : ``33.xxxxxx", ``lng" : ``-118.xxxxxx"\}} \\ \hline
    Take down light Christmas decorations & \small{``radius" : ``0.5", ``geo\_center" : \{``lat" : ``33.xxxxxx", ``lng" : ``-118.xxxxxx"\}} \\ \hline
    Put up 20 yard sale signs in Mid-Wilshire area & \small{``radius" : "0.5", ``geo\_center" : \{``lat" : ``33.xxxxxx", ``lng" : ``-118.xxxxxx"\}} \\ \hline
    \end{tabular}
    \label{tab:requested_tasks}
    \end{center}
\end{table}

\section{Privacy Countermeasures}
\label{sec:3}

In this section we survey some state-of-the-art approaches addressing the privacy issues in spatial crowdsourcing. We first categorize the studies into two groups: \emph{tasking in the push mode} and \emph{reporting in the pull mode}. Subsequently, each subgroup is further classified according to the applied techniques. Within each subgroup we identify one key paper shown in boldface to be presented in depth while follow-up studies are briefly discussed. An overview of these studies is presented in Table~\ref{tab:papers}. The table shows that the studies solely focus on location privacy of workers and assume that the locations and content of tasks are public. Moreover, the SC-server is regarded as a primary threat in all studies, while some consider workers and requesters as secondary adversaries. We also notice that the most recent studies focus on the push mode, which requires privacy protection during tasking. This problem is considerably more challenging when compared to the problem of privacy-preserving reporting in the pull mode.

\begin{table*}
\begin{center}
\caption{Overview of problem focuses (Re: reporting, Ta: tasking); privacy techniques used (Ps: pseudonym, Cl: cloaking, Pt: perturbation, Ex: exchange-based, En: encryption-based); threats (W: worker, T: requester, S: server); trusted third party (TTP); optimization type (ST: single task, MT: multiple tasks). x and (x) represent primary and secondary aspects, respectively.}
\label{tab:papers}
\begin{tabular}{l | c  c | c c  c  c  c   | c c | c  c  c | c  c | c c}
\multirow{2}{*}{Paper} & \multicolumn{2}{ c |}{Phase} & \multicolumn{5}{ c |}{Techniques} & \multicolumn{2}{ c |}{Protection} & \multicolumn{3}{ c |}{Threats} & \multicolumn{2}{ c |}{TTP}  & \multicolumn{2}{ c }{Opt. type} \tn\cline{2-17}
\multicolumn{1}{c |}{} & \centering Re & \centering Ta & \centering Ps & \centering Cl & \centering Pt & \centering En & \centering Ex  & \centering W & T \centering & \centering W & \centering R & \centering S  & \centering Yes & \centering No & \centering ST & \centering MT \tn
\hline
\textbf{Shin et al. 2011}~\cite{Shin2011}  											& x & x		& x & (x) & & (x) &       &  x & N/A	   	& & N/A & x 		  	 & x &  	 		& x &  \tn 
\hline
\textbf{Boutsis et al. 2013}~\cite{Boutsis2013}  								& x & 		& (x) & & &  & x	    & x & N/A   	& (x) & N/A & x 				&  & x 		 		& & x  \tn 
\hline
Zhanget al. 2016~\cite{Zhang2016} 									& x & 		& &  & &  & x 	  &  x & N/A    	&  & N/A & x &  	         	& x			& & x  \tn 
\hline
\textbf{Kazemi et al. 2011}~\cite{Kazemi2011a}  							&  &	x		& (x) & x & & &      & x &   		& & (x) & x 		  		& x & 			&  & x \tn 
\hline
Vu et al. 2012~\cite{Vu2012}  											& & x 		&  & x & & (x) & 	    & x &     	& (x) & (x) & x 				& x & 			& & x \tn 
\hline
\textbf{To et al. 2014}~\cite{To2014} 							&  & x 			& &  & x & &		     & x &      	& (x) & (x) & x 			 & x & 	 		& x & \tn 
\hline
Gong et al. 2015~\cite{Gong2015} 						&  & x  			& &  & x & &		    & x &   		& (x) & (x) & x 				& x & 	 		& x & \tn
\hline
Zhang et al. 2015~\cite{Zhang2015} 						&  & x			& & & x & &		    & x &    		& (x) & (x) & x 				& x &  	 		& x &  \tn 
\hline
To et al. 2016~\cite{To2016b} 				&  &	x		& &  & x & & 	  &  x &    	& (x) & (x) & x 		    	  	& x  & 		 			& x & \tn 
\hline
\textbf{Pournajaf et al. 124}~\cite{Pournajaf2014} 					&  & x		& & x & & &		  &  x &    	& & (x) & x 			& x &  	 		& & x \tn 
\hline
Hu et al. 2015~\cite{Hu2015} 										&  & x			& & x & & &		    & x &     	& & (x) & x 				& x &   	 		& & x \tn 
\hline
Shen et al. 2916~\cite{Shen2016} 								&  & x			& & & & x &		   & x &    & (x) & (x) & x 							&   & x	 		& x &   \tn 
\end{tabular}
\end{center}
\end{table*}

\subsection{Protection in the Pull Mode}

Privacy protection in the pull mode has been studied in the context of participatory sensing.
In this section we highlight recent studies that often focus on the reporting phase of the pull mode. They use either \emph{pseudonymity}~\cite{Shin2011} or \emph{exchange-based} techniques~\cite{Boutsis2013,Zhang2016}. The pseudonymity method disassociates the connections between one's uploaded data and his/her identity while the latter exchanges workers' crowdsourced data and location information before uploading them to a server so that the server is uncertain about locations of individual workers.

\subsubsection{Pseudonymity Techniques}

\begin{figure}[ht]
		\centering
		\includegraphics[width=0.8\textwidth]{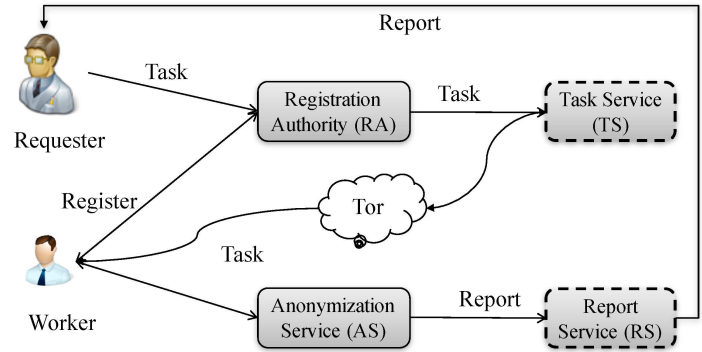}
		\caption{A framework for privacy protection during tasking and reporting in the pull mode. Dashed entities are malicious, while others are trusted.}
		\label{fig:anonysense}
\end{figure}

Shin et al.~\cite{Shin2011} propose a privacy-preserving framework for the pull mode as illustrated in Figure~\ref{fig:anonysense}.
A requester submits a task to a \emph{registration authority} (RA) that will verify the task before sending it to a \emph{task service} (TS). Also, a worker connects to TS through an anonymizing network such as Tor to request new tasks, referred to as a \emph{task subset}.
After receiving the requested tasks, the worker chooses which tasks to accept. He then performs the tasks and uploads the corresponding task reports to a \emph{report service} (RS) via an \emph{anonymous service} (AS). In this framework, RA and AS are trusted while TS, RS and requesters can be hostile. TS and RS can be considered as services performed by the SC-server.

This study~\cite{Shin2011} provides privacy protection in both tasking and reporting phases. During tasking, the role of the anonymizing network is to disassociate the worker and his requested tasks, depicted by the first and the third assignment links in Figure~\ref{fig:threat_model_pull}.
To preserve privacy during reporting, a worker typically sends his task report to RS via AS, which routes the report through multiple servers so that the SC-server (i.e., TS and RS) cannot associate multiple locations (i.e., IP addresses) with the identity of the same worker. Consequently, the SC-server is oblivious to the first reporting link in Figure~\ref{fig:threat_model_pull}.
More recently, there has been closely related work in participatory sensing that enables workers to hide their locations and data ownership by passing the collected data through a random neighboring worker multiple times before uploading the data to the SC-server~\cite{Hu2010}.

\subsubsection{Exchange-Based Techniques}

Pseudonymity techniques are ad hoc and do not provide quantifiable privacy protection.
For more sensitive tasks that require strong privacy guarantee, $k$-anonymity~\cite{sweeney2002k} is used in~\cite{Shin2011} to ensure that each report is anonymized with $k-1$ reports generated by other workers with similar sensitive information. However, such techniques may not be applicable to SC because the worker location is part of the report. To address such a problem, Boutsis and Kalogeraki~\cite{Boutsis2013} propose the exchange-based technique to obscure the workers by exchanging their reports between them before disclosing the sensitive information to an untrusted server (i.e., SC-server). Such a technique can be used as AS in Figure~\ref{fig:anonysense}, aiming to protect the first reporting link in Figure~\ref{fig:threat_model_pull} from the SC-server.

To provide a quantifiable privacy guarantee, in~\cite{Boutsis2013} the authors use location entropy as the measure of privacy or the attacker's uncertainty. The study aims to make all workers' trajectories as equiprobable to contain sensitive locations by maximizing the location entropy of an individual's trajectories to be defined later. To maximize the location entropy, trajectories with sensitive locations are distributed among multiple workers. Particularly, each worker's mobile phone identifies the $k$ most frequently visited locations as \emph{sensitive data} from a \emph{local} trajectory database. A trajectory is selected for exchange if removing the trajectory increases the entropy of the database, computed as follows.
$$H_i=\sum_{loc_{ij}\in L}Pr(loc_{ij})log(Pr(loc_{ij})))$$
where $L$ is the set of locations and $Pr(loc_{ij})$ is the fraction of total visits to location $j$ that belongs to user $i$. Consequently, an attacker will not be able to identify sensitive locations or identities of the workers.

For each worker, the trajectories that contain locations with high frequency are exchanged with other workers since removing high-frequency trajectories (trajectories that contain sensitive locations) makes the frequency of the locations in $L$ more homogeneous and thus increases the entropy. Furthermore, as other workers may not be trustful, not only the set of high-frequency trajectories are exchanged but also another set of trajectories that do not contain the sensitive locations. This guarantees that neighboring workers are not able to associate the worker with his sensitive data. Consequently, both frequent and non-frequent trajectories are selected and forwarded to individual workers so that no worker can be certain about the sensitivity of any trajectory.

A drawback of computing entropy locally is that the exchange decisions can be suboptimal due to the lack of a global view of all workers. This is because individual workers try to maximize their own entropy regardless of each other, which goes contrary to the fact that exchanging trajectories alters the location entropy of multiple workers. Thus, the exchange-based technique should consider the entropy with respect to all workers as opposed to individual workers. Therefore, Zhang et al.~\cite{Zhang2016} introduce a similar framework, but here workers coordinate with each other to exchange their sensing data, including locations before uploading to the SC-server. As a result, all sensitive locations are equally likely visited by any worker so that the actual trajectory of each worker cannot be learned. However, unlike~\cite{Boutsis2013} where entropy is computed for a single worker, here entropy is calculated for all workers.

Although the exchange-based technique is simple and does not rely on a trusted server, the actual location information is still uploaded to the SC-server. Therefore, this approach is vulnerable to background knowledge attack. For instance, if the SC-server knows that only worker $w_i$ visits a particular location where a report was uploaded, the server is certain that $w_i$ actually made the report. 


\subsection{Protection in the Push Mode}
While preserving privacy during reporting in the pull mode has been largely studied in the context of participatory sensing (a recent survey can be found in~\cite{Christin2016a}), recent SC studies focus on the more challenging phase of tasking. These studies generally assume the push mode. 
We emphasize that focusing on the tasking step in the push mode is the correct approach, given that SC workers have to physically travel to the task location. The completion of a task discloses the fact that some worker must have been at that location, and this is unavoidable in SC. 
Focusing on tasking also makes sense from a disclosure volume standpoint. During the assignment, all workers are candidates for participation; therefore, locations of all workers are exposed, absent a privacy-preserving mechanism. Nevertheless, after task request dissemination, only a few workers will participate in task completion, and {\em only if they give their explicit consent} (see the threat model for the push mode in Section~\ref{sec:threat_model}).

Various techniques have been proposed to protect location privacy of workers during task assignment in SC, including \emph{cloaking} (hide the accurate location in a cloaked region)~\cite{Kazemi2012,Vu2012,Pournajaf2014,Hu2015}, \emph{perturbation} (distort the actual location information by adding artificial noise)~\cite{To2014,Gong2015,Zhang2015,To2016b} and \emph{encryption}~\cite{Shin2011,Shen2016}.

\subsubsection{Cloaking Techniques}

The studies in this category generally implement spatial $k$-anonymity by generating a \underline{c}loaking \underline{r}egion (CR) for each worker, which includes $k-1$ other workers. To guarantee strong privacy protection, peer-to-peer spatial $k$-anonymity~\cite{chow2011spatial} has been adopted in these studies. In the following we first present a simplified version of tasking without constraints. Next, we survey some recent studies that consider real-world constraints, such as the travel budget of each worker and a worker's willingness to perform tasks.

\paragraph{\textbf{Task Assignment Without Constraints}}

In~\cite{Shin2011}, each worker requests a \emph{task subset} of size $p$ at a time; however, choosing an appropriate value of $p$ is not trivial. Large $p$ may lead to not only high communication overhead between workers and TS, but tasks are also \emph{unnecessarily} disclosed to the workers. In contrast, small $p$ may result in some tasks that will never be accepted by any worker. One reason for this is that a worker can browse far-away tasks that he cannot complete before the tasks' deadlines. This redundant disclosure incurs additional privacy threats to the requesters of those tasks.

In order to minimize such disclosure, Kazemi and Shahabi~\cite{Kazemi2011a} propose a privacy framework that enables each worker $w_i$ to query the SC-server for a set of nearby spatial tasks. Particularly, the SC-server needs to distribute a set of spatial tasks to workers such that each worker is assigned a subset of tasks that are closer to him than to any other worker.  Without privacy protection, the SC-server can construct a Voronoi diagram of the workers, including a set of cells where each cell belongs to a worker, and any spatial task in the cell is closer to the worker than to any other worker. Once the server computes the Voronoi diagram of the workers, it forwards to each worker all the spatial tasks lying inside the corresponding cell. However, in such a scenario, an adversary may infer the worker's identity by associating the query to query location (i.e., the location from which the query is issued. This is referred to as \emph{location-based attack}. Consequently, the framework aims to protect worker identity from location-based attacks by disassociating a query from the query location\footnote{However, this study assumes that workers trust one another. Hence, a more recent study~\cite{Vu2012} solves a similar problem as in~\cite{Kazemi2011a} without the assumption of trusted workers.}. The framework named PiRi (partial-inclusivity and range independence) has both  \emph{query formation} and \emph{query selection}.

\begin{figure*}[ht]
	\begin{minipage}[b]{0.305\linewidth}
		\centering
		\includegraphics[width=1\textwidth]{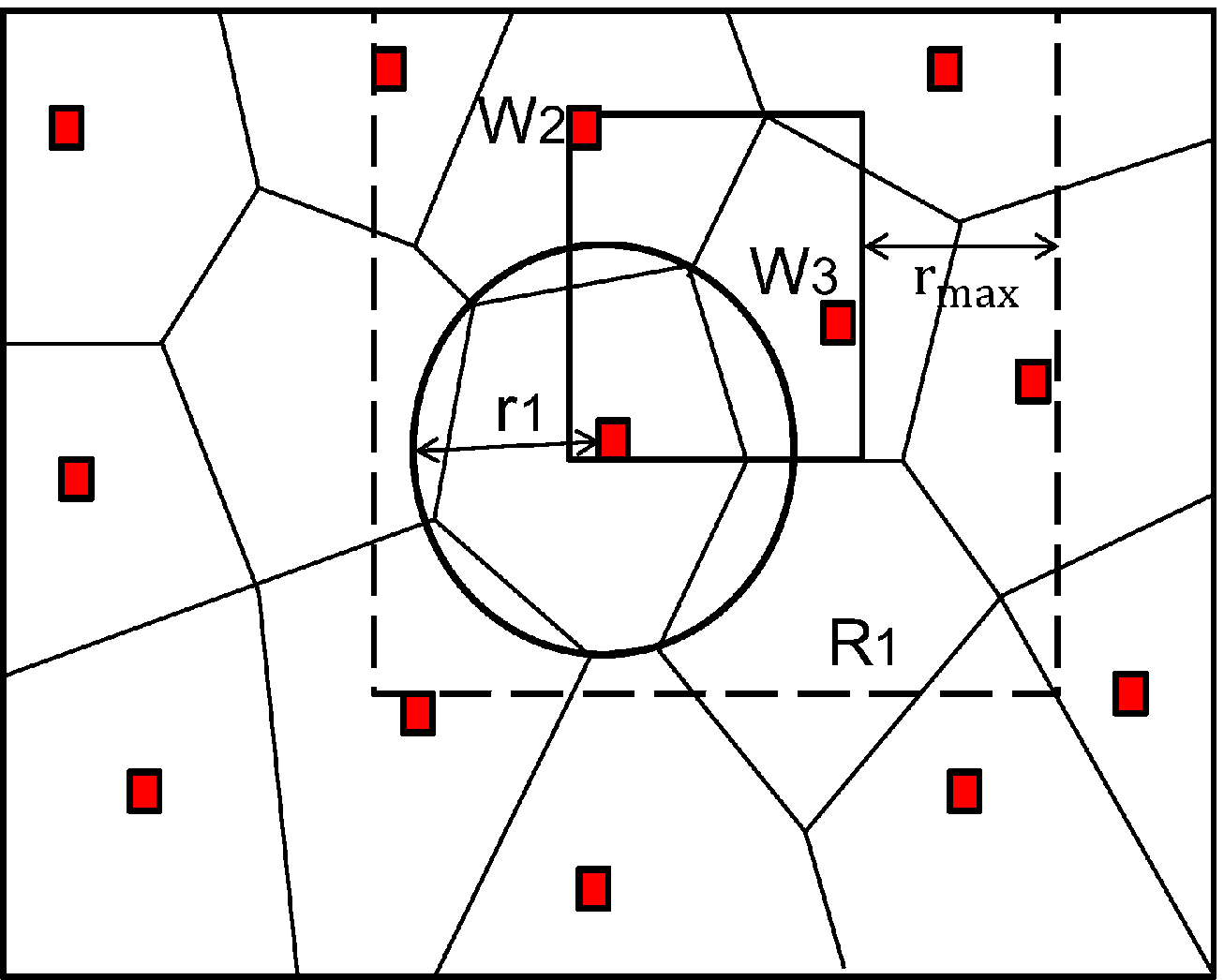}
		\subcaption{Query formation}
		\label{fig:query_formation}
	\end{minipage}
	\vspace{1pt}
	\begin{minipage}[b]{0.305\linewidth}
		\centering
		\includegraphics[width=1\textwidth]{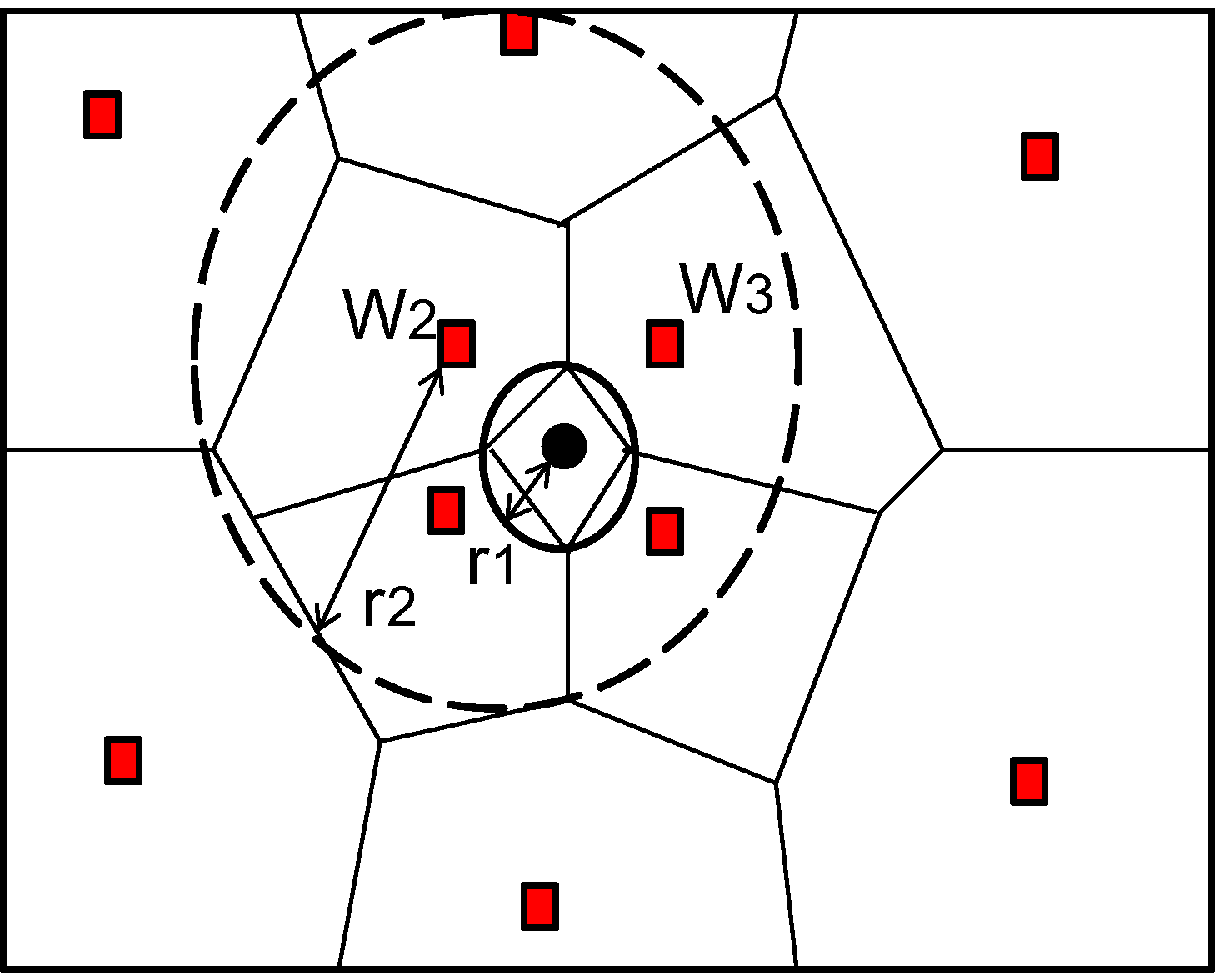}
		\subcaption{Range dependency leak}
		\label{fig:range_dependency}
	\end{minipage}
	\vspace{1pt}
	\begin{minipage}[b]{.37\linewidth}
		\centering
		\includegraphics[width=1\textwidth]{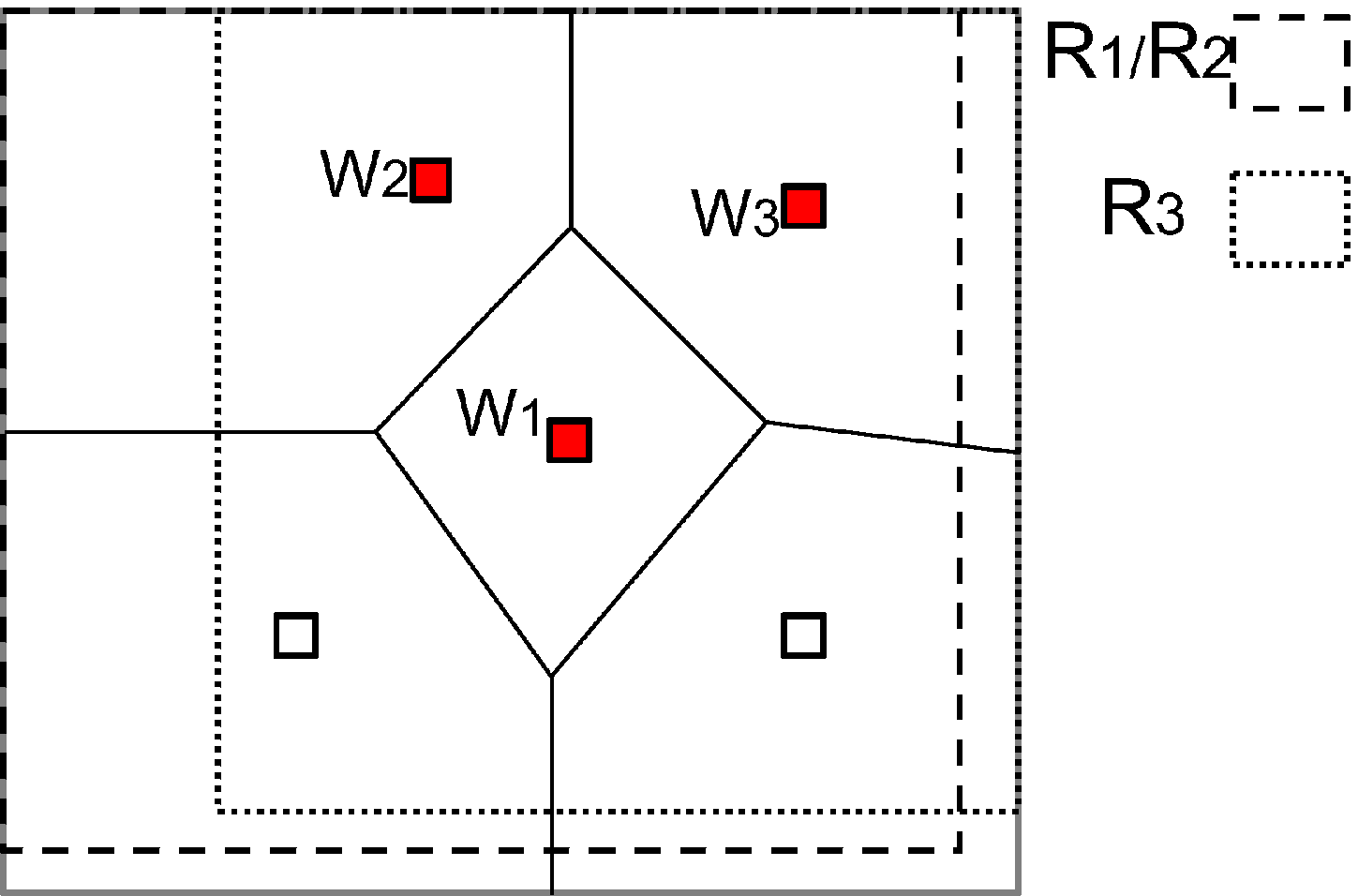}
		\subcaption{All-inclusivity leak}
		\label{fig:all_inclusivity}
	\end{minipage}
	\caption{Examples of range dependency and all-inclusivity.}
	\label{fig:piri}
\end{figure*}

\subparagraph{Query Formation}
In the query formation step, each worker $w_i$ computes his Voronoi cell by communicating with his neighboring peers~\cite{chow2011spatial}. The worker forms his CR, where his location is blurred among $k-1$ other peers (with $k=3$, the solid-lined rectangle in Figure~\ref{fig:query_formation}). The worker can send the CR along with the radius $r$ (i.e., the smallest enclosing circle of $w_i$'s Voronoi cell) to the server to retrieve all the tasks which lay inside his Voronoi cell. However, the range query is dependent on the size of the worker's Voronoi cell (range dependency), which is a potential for information leaks. Considering an extreme scenario where the server knows the workers' locations, it also knows their Voronoi cells and therefore the radius $r$ for each of them. Consequently, the server can easily identify the query issuer (i.e, the set of all workers in the CR with radius $r$). Figure~\ref{fig:range_dependency} depicts such a scenario, where $w_1$ (black-filled circle) cloaks himself with $w_2$, and sends the CR along with radius $r_1$ to the server (see the size of $r_1$ as compared to $r_2$). The server, knowing the location of the workers, and hence their Voronoi cells (i.e., $r_1$, and $r_2$), relates the query with radius $r_1$ to its query location (i.e., the location of a worker with the Voronoi cell of the same radius).

In order to avoid the range dependency leak, each worker $w_i$ should cloak not only his location but also his range query among $k-1$ other peers. In other words, instead of forming his range query with radius $r_i$, the worker forms his query with radius $r_{max}$--the maximum radius among all the $k$ peers inside the CR. This guarantees the $k$-anonymity at all times. In Figure~\ref{fig:query_formation}, $R_1$ (the dotted line rectangle) shows the query region formed by $r_{max}$.

\subparagraph{Query Selection}
Once all workers have formed their query regions, they can send them out to the server. However, the SC-server can utilize the gathered information (i.e, query regions) from all workers to attack the system (all-inclusivity leak). Figure~\ref{fig:all_inclusivity} illustrates such scenario, in which workers $w_{1..3}$ participate in the system. The figure shows that $w_1$ cloaks himself with $w_2$. Similarly, $w_2$ forms a cloaked region with $w_1$. Subsequently, both $w_1$ and $w_2$ form identical query regions. The figure also depicts that $w_3$ cloaks himself with $w_1$. Accordingly, the server can easily identify $w_3$ by relating it to the query region $R_3$, since $w_3$ appears only once (i.e., $R_3$) in all the three submitted query regions to the server. This indicates that the more workers submit queries to the server, the more information the server has to infer the workers' identities. To prevent this leak, the authors attempt to minimize the number of queries submitted to the server while assigning the nearby tasks to \emph{every} single worker.

Since there is a large overlap among the query regions of the workers, a worker can share his result received from the server with all the peers whose Voronoi cells lay completely inside his query region. The problem is how to select the group of representative workers, formally stated as follows.
Given a set of workers $W$, and a set of spatial tasks $T$, let $R$ and $V$ be the set of query regions and Voronoi cells for the set $W$, respectively, where $R_i$ corresponds to the query region for worker $w_i$, and $V_i$ is the Voronoi cell for $w_i$. The problem is to find a set $C\subseteq R$ that covers the entire set $V$ with minimum cardinality. This problem is shown to be NP-hard by reduction from the minimum set cover problem~\cite{Kazemi2011a}. One well-known approach for solving the set cover problem is a greedy algorithm that picks a representative worker whose query region covers the largest number of uncovered Voronoi cells from $V$. However, this approach is applicable only in a centralized setting, where a global knowledge of the environment is available.
To address this issue, the greedy heuristic is extended to support the distributed environment. Particularly, a voting mechanism is devised to select the set of representative workers, whose CRs are sent out to the server. These query results will later be shared with the rest of the workers. This step has been shown to prevent the all-inclusivity leak~\cite{Kazemi2011a}.

\paragraph{\textbf{Task Assignment with Constraints}}

In~\cite{Kazemi2011a,Vu2012}, spatial tasks are distributed to the corresponding nearest workers. This objective may not necessarily fit SC applications as workers often have various constraints that need to be considered. For example, they may be willing to perform  tasks that are far away, but within their daily travel routes. To capture such constraints, each worker $w_i$ has a cloaked area $a_i$ and a limited travel budget $b_i$, which denotes the maximum distance he is willing to travel~\cite{Pournajaf2014}. Given the cloaking regions of a set of workers, the objective of the SC-server is to match a set of spatial tasks to the workers such that task assignment is maximized while satisfying the travel budget constraint of each worker.

As \emph{travel cost} (often measured by the distance between tasks and assigned workers) is an important performance metric in SC, in the following we first present two methods for estimating the travel cost from the cloaked areas of the workers. Thereafter, we present the problem of \underline{s}patial \underline{t}ask \underline{a}ssignment with \underline{c}loaked locations (STAC)~\cite{Pournajaf2014}.

\subparagraph{Distance Estimation}

\begin{figure*}[ht]
\hspace{5pt}
	\begin{minipage}[b]{0.33\linewidth}
		\centering
		\includegraphics[width=1\textwidth]{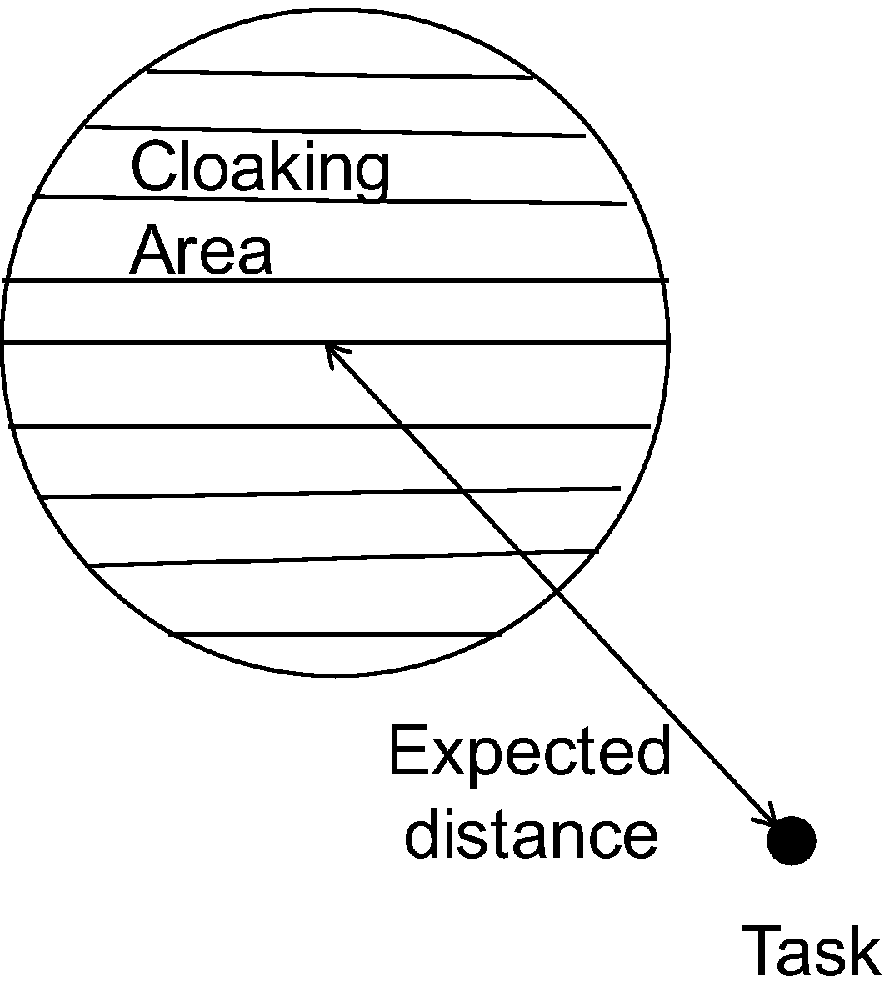}
		\subcaption{Centroid-point method}
		\label{fig:centroid_point}
	\end{minipage}
	\hspace{10pt}
	\begin{minipage}[b]{.45\linewidth}
		\centering
		\includegraphics[width=1\textwidth]{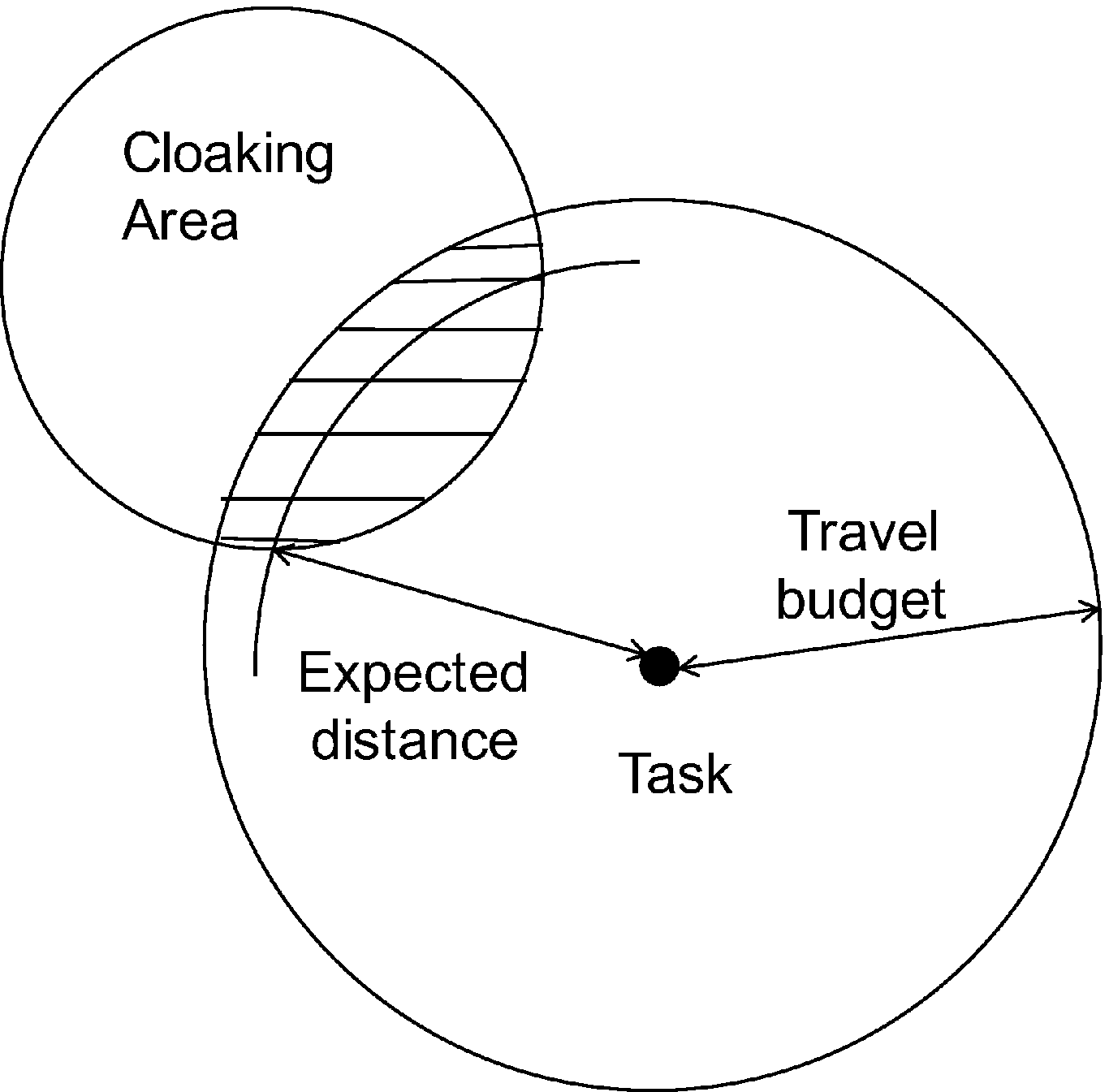}
		\subcaption{Expected probabilistic method}
		\label{fig:expected_probabilistic}
	\end{minipage}
	\caption{Distance estimation methods.}
	\label{fig:distance_estimation}
\end{figure*}

Given the cloaked area $a_i$ of the workers, STAC proposes two methods for estimating the expected distances between pairs of workers $w_i$ and tasks $t_j$, named $\hat{d}_{i,j}$. The baseline method approximates the worker location as the centroid of his cloaking area as depicted in Figure~\ref{fig:centroid_point}. Another method uses the travel budget of the worker to prune the cloaking area (i.e., the dashed area in Figure~\ref{fig:expected_probabilistic}), resulting in a shrunk area that contains only accessible locations of the worker. Consequently, $\hat{d}_{i,j}$ is estimated by the distance between the task location and shrunk areas.

Next, we present a two-phase optimization approach to STAC. The first phase, denoted as G-STAC, globally optimizes task assignment using cloaked locations of the workers. The second phase, referred to as L-STAC, locally optimizes the assignment of individual workers using their own exact locations.

\subparagraph{Global Optimization}

Given a set of workers and a set of spatial tasks, G-STAC aims to achieve a particular goal of task coverage with the minimum travel cost. G-STAC is formally defined as follows.

\begin{displaymath}
\begin{split}
& min\ TC = \sum\limits_{i\in W} \sum\limits_{j\in T}\hat{d}_{i,j} x_{i,j} \\
& s.t.\ TU = \sum\limits_{j\in T}\frac{\sum_{i\in W} x_{i,j}}{k_j} \ge gm \\
& \sum\limits_{j \in T}\hat{d}_{i,j} x_{i,j} \le b_i \\
\end{split}
\end{displaymath}
where task cost ($TC$) is the total distance traveled by all workers, while task coverage or utility ($TU$) is the total covered fraction of tasks. $\hat{d}_{i,j}$ is the estimated distance between worker $i$ and task $j$, $x_{i,j}=1$ means worker $i$ is assigned to task $j$, otherwise $x_{i,j}=0$, $k_j$ is the required coverage of $t_j$ (i.e., the number of workers to perform $t_j$) while $g\in (0,1]$ indicates the required fraction of  coverage for a task. The last constraint guarantees that $w_i$'s travel distance is within his budget $b_i$.

G-STAC is shown to be NP-hard by reduction from the minimum set cover problem. Therefore, a greedy algorithm is proposed that iteratively selects the most cost-effective worker-task pair and updates $TU$ until either the coverage goal is achieved or the travel budgets of all workers are spent. A worker-task pair is cost-effective if the ratio of expected distance to the expected coverage contributed by this worker is small.

\subparagraph{Local Optimization}
The output of G-STAC is the best mapping of tasks to workers, which is sent to workers as suggested assignments. However, a worker may be assigned tasks whose locations exceed his travel budget, or nearby tasks are not assigned to him because their distance has been estimated as being farther away. Thus, the local refinement phase (L-STAC) is performed by individual workers' devices for more coverage and lower travel cost. A caveat is that selecting the closest tasks for each worker may result in over-coverage for some tasks, while the others remain unperformed. Consequently, in addition to minimizing the travel cost, L-STAC also tries to minimize the change in the local optimization when compared to the global optimization. L-STAC is formally defined as follows.

\begin{displaymath}
\begin{split}
& min\ TC_i = \sum\limits_{j\in T}d_{i,j} y_{i,j} \\
& s.t.\ |y_i - x_i| < \epsilon \\
& \sum\limits_{j\in T}\frac{y_{i,j}}{k_j} \ge \sum\limits_{j\in T}\frac{x_{i,j}}{k_j}  \\
& \sum\limits_{j \in T}d_{i,j} y_{i,j} \le b_i \\
\end{split}
\end{displaymath}
where for each worker $w_i$, $x_i$ and $y_i$ are the binary assignment vectors of the global and local phases of STAC, respectively. The first constraint, $|y_i - x_i|$, is the Hamming distance between $x_i$ and $y_i$, which is bounded by a threshold $\epsilon$ aiming to keep minimum changes in the local assignment. The second constraint ensures that $w_i$'s contribution to the task coverage is not decreased when compared to his contribution in the global phase. In the same fashion, L-STAC is NP-hard by reduction from the minimum set cover problem; thus, another greedy algorithm has been proposed to solve L-STAC.



Recently, Hu et al.~\cite{Hu2015} extended the travel budget constraint in~\cite{Pournajaf2014} to a \emph{spatial region}, represented by a rectangle $R$, within which the worker is willing to travel. Similar to~\cite{Kazemi2011a}, workers employ the peer-to-peer cloaking technique~\cite{chow2011spatial} to cloak their locations among $k-1$ other workers. Also, each worker's cloaking area must contain his spatial region $R$, otherwise the cloaking area is extended to cover $R$. Observing that workers' cloaking areas often contain multiple spatial regions of other workers, to reduce the communication overhead, only some cloaking areas that could cover all the workers' spatial regions will be sent to the SC-server. This technique limits the disclosure of information when compared to sending all the workers' cloaking areas to the SC-server~\cite{Pournajaf2014}.

The cloaking techniques used in~\cite{Pournajaf2014,Hu2015} are intuitive; nevertheless, their privacy guarantee is weak. Such obfuscation-based techniques do not provide rigorous privacy protection and are prone to homogeneity attack~\cite{machanavajjhala2007diversity} when all $k$ workers are at the same location.  Also, the value $k$ needs to be specified to guarantee the desired level of privacy protection. Unfortunately, choosing an appropriate $k$ value can be difficult because $k$-anonymity does not consider the frequency of user visits. To elaborate, a location may be visited by many workers---those who have a dominant contribution to the location (i.e., home or office) are most likely to be the subject of attack. Consequently, one with a background knowledge of who visits the location the most can easily perform such an attack.

\subsubsection{Perturbation Techniques}

Methods in this category use differential privacy (DP) to protect workers' locations during task assignment~\cite{To2014,To2015,Gong2015,Zhang2015,To2016b}, which overcomes the aforementioned issues of the obfuscation technique.
DP has emerged as the de facto standard with strong protection guarantees rooted in statistical analysis. It provides a \emph{semantic} privacy model as opposed to a \emph{syntatic} model in other sanitization techniques (e.g., $k$-anonymity, $l$-diversity). DP has been adapted by major industries for various tasks without compromising individual privacy, e.g., discovering users' usage patterns with Apple~\cite{appledp2016} or crowdsourcing statistics from end-user client software with Google~\cite{erlingsson2014rappor}. DP ensures that an adversary is not able to reliably learn from the published sanitized data whether or not a particular individual is present in the original data, regardless of the adversary's prior knowledge.

The authors in~\cite{To2014} propose system model, privacy model and performance metrics, followed by two main steps that preserve privacy and identity of workers: \emph{sanitization} of workers' locations and \emph{task assignment} on the sanitized data.

\paragraph{\textbf{System Model}}

\begin{figure*}[ht]
	\begin{minipage}[b]{0.50\linewidth}
		\centering
		\includegraphics[width=1\textwidth]{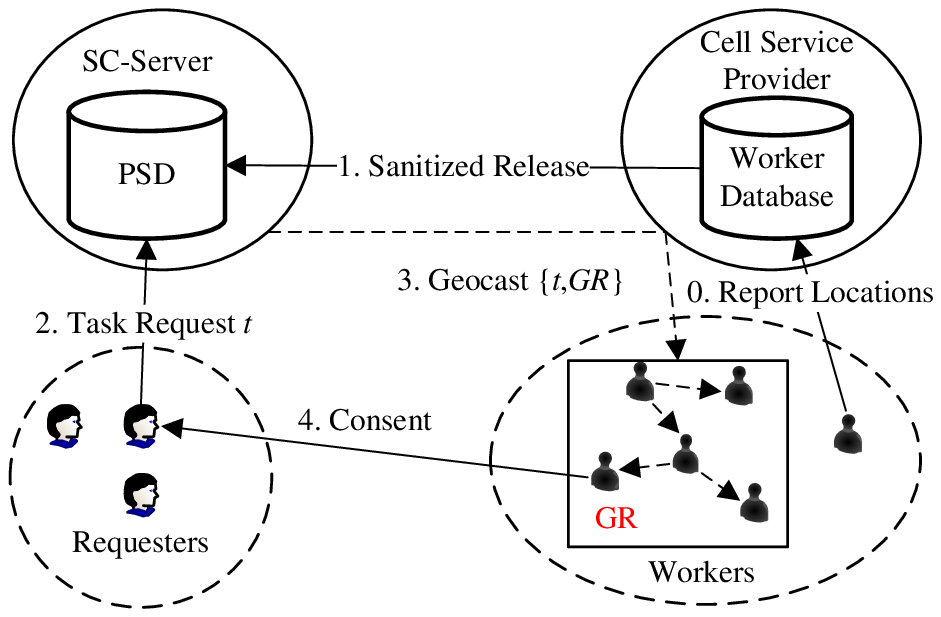}
		\subcaption{System architecture}
		\label{fig:framework}
	\end{minipage}
	\hspace{5pt}
	\begin{minipage}[b]{.45\linewidth}
		\centering
		\includegraphics[width=1\textwidth]{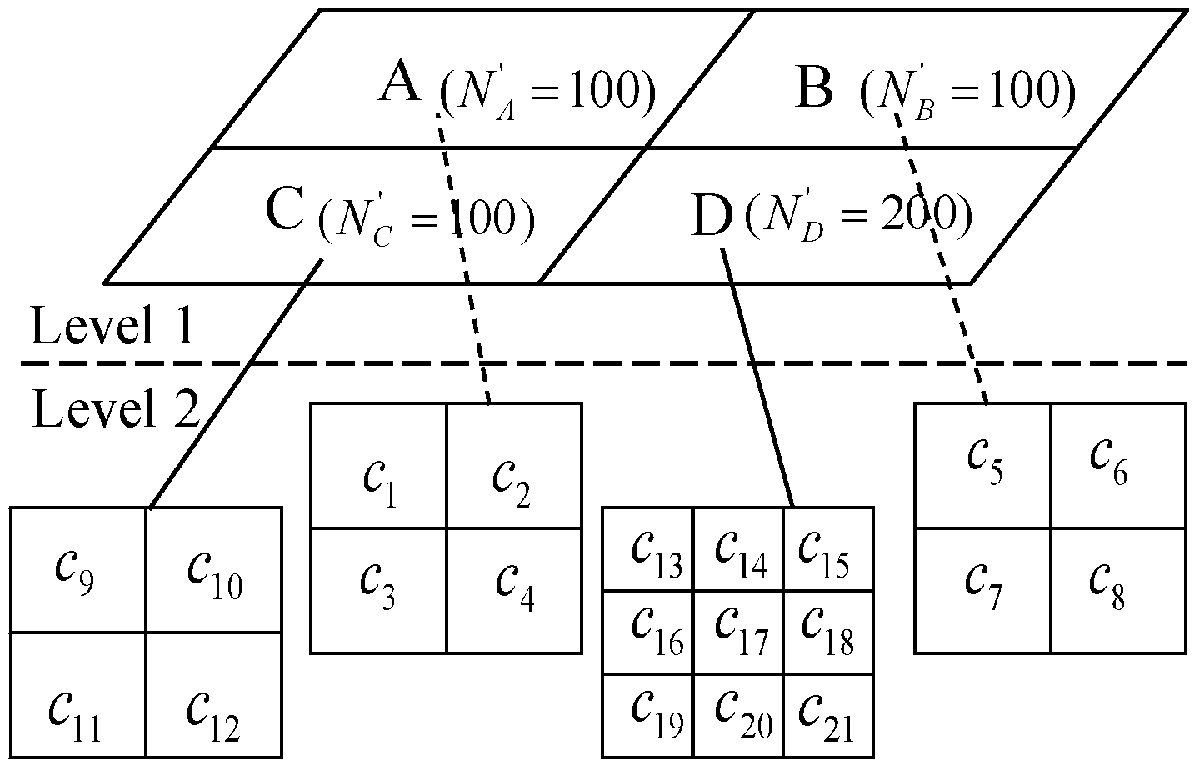}
		\subcaption{Worker PSD using adaptive grid}
		\label{fig:AG}
	\end{minipage}
	\caption{Differentially private framework for spatial crowdsourcing.}
	\label{fig:ccc}
    \vspace{-10pt}
\end{figure*}

To protect location privacy of workers participating in spatial tasks, the SC-server must only have access to data sanitized according to {\em $\epsilon$-differential privacy}~\cite{dwork2006differential} ($\epsilon$ is privacy loss or privacy budget). 
Figure~\ref{fig:framework} shows the system architecture. Workers send their locations (Step $0$) to a trusted {\em cellular service provider} (CSP) which collects updates and releases a \emph{private spatial decomposition} (PSD) according to privacy budget $\epsilon$ mutually agreed upon with the workers. The PSD is accessed by the SC-server (Step $1$), which also receives tasks from a number of requesters (Step $2$).
When the SC-server receives a task $t$, it queries the PSD to determine a {\em geocast region (GR)} that encloses with high probability workers close to $t$. Next, the SC-server initiates a {\em geocast} communication \cite{navas1997geocast} process (Step $3$) to disseminate $t$ to all workers within $\mathit{GR}$. According to DP, sanitizing a dataset requires creation of fake locations in the PSD. If the SC-server is allowed to directly contact workers, then failure to establish a communication channel would breach privacy, as the SC-server is able to distinguish fake workers from real ones. Using geocast is a unique feature of the framework which is necessary to achieve privacy protection. Geocast can be performed either with the help of the CSP infrastructure, or through a mobile ad hoc network where the CSP contacts a single worker in the $\mathit{GR}$, and then the message is disseminated on a hop-by-hop basis to the entire $\mathit{GR}$. The latter approach keeps CSP overhead low and can reduce operation costs for workers. 
Upon receiving request $t$, a worker $w_i$ decides whether to perform the task or not. If yes (Step $4$), she sends a {\em consent} message to the SC-server (or requesters) confirming $w_i$'s availability. 
If $w_i$ is not willing to participate in the task, then no consent is sent, and no information about the worker is disclosed.


\paragraph{\textbf{Privacy Model and Assumptions}}

The objective of the framework is to protect both the {\em location} and the {\em identity} of workers {\em during task assignment}. Once a worker consents to a task, the worker herself may directly disclose information to the task requester (e.g., to enable a communication channel between worker and requester). However, such additional disclosure is outside the scope of this work, as each worker has the right to disclose his or her individual information. Instead, the focus of the framework is on what happens prior to consent, when worker location and identity must be protected from {\em both} task requesters and the SC-server. This privacy model is a weaker version of the restrictive model in Figure~\ref{fig:threat_model_push} since task locations are public.

Workers cannot trust the SC-server, especially as there may be many such entities with diverse backgrounds, e.g., private companies, non-profits, government organizations, academic institutions. On the other hand, the CSP already has a signed agreement with workers through the service contract, so there is already a trust relationship established, as well as mutually-agreed upon rules for data disclosure. Furthermore, the CSP already knows where subscribers are, e.g., using cell tower triangulation, so worker location reporting does not introduce additional disclosure. 
In addition, having the CSP expose a PSD release of the user location dataset can benefit applications beyond crowdsourcing. For instance, the PSD can be shared with law enforcement agencies for public safety, or with commercial organizations to increase the revenue of the CSP. Therefore, there is sufficient motivation for the CSP to provide such a location sanitization service.

However, the CSP has no expertise, and perhaps no financial interest, to host an SC service, which needs to deal with a diverse set of issues such as interacting with various task requester categories, managing profiles (e.g., some workers may only volunteer for environmental tasks), etc. The role of the CSP is to aggregate locations from subscribed workers, transform them according to DP, and release the data in sanitized form to one or more SC-servers for assignment. As multiple SC-servers can use the same PSD, it is practical for the CSP to provide PSDs for a small fee, e.g., a percentage of the workers' payment, or a tax incentive in the case of a public-interest SC application.

\paragraph{\textbf{Design Goals and Performance Metrics}}

Protecting worker location significantly complicates task assignment and may reduce the effectiveness and efficiency of worker-task matching. Due to the nature of DP, it is possible for a region to contain no workers, even if the PSD shows a positive count. Therefore, no workers (or an insufficient number thereof) may be notified of the task request, and the task may not be completed.
Alternatively, the GR may comprise workers who are a long distance away from the task location, whereas nearer workers are not included.
Finally, in the non-private case, only one selected worker, whose location and identity is known, is notified of the task request. With location protection, redundant messages need to be sent, increasing overhead.
We focus on the following performance metrics:

\begin{itemize}
\item
\emph{Assignment success rate (ASR).} Due to PSD data uncertainty, the SC-server may incorrectly assign workers to tasks (e.g., no worker is reached, or task is too far and workers do not accept it). $\mathit{ASR}$ measures the ratio of tasks accepted by a worker
to the total number of task requests.
\item
\emph{Worker travel distance (WTD).} The SC-server is no longer able to accurately evaluate worker-task distance, hence workers may have to travel long distances to tasks. The challenge is to keep the worker travel distance low, even when exact worker locations are not known.
\item
\emph{System overhead.} Dealing with imprecise locations increases the complexity of assignment, which poses scalability problems. 
A significant metric to measure overhead is the {\underline a}verage number of {\underline n}otified {\underline w}orkers ($\mathit{ANW}$).
This number affects both the {\em communication overhead} required to geocast task requests, as well as the {\em computation overhead} of the matching algorithm, which depends on how many workers need to be notified of a task request.
\end{itemize}

\paragraph{\textbf{Sanitization of Workers' Locations Using Adaptive Grid}}

The first step in the proposed framework consists of building a PSD (at the CSP side) to be used later for task assignment at the SC-server. Building the PSD is an essential step because it determines how accurate  the released data is, which in turn affects $\mathit{ASR}$, $\mathit{WTD}$ and $\mathit{ANW}$.
Worker location data are sanitized at the CSP using a PSD, named \emph{adaptive grid} (AG)~\cite{qardaji2013differentially}. PSD is a sanitized spatial index, where each index node contains a noisy count of the workers rooted at that node. Figure~\ref{fig:AG} shows a snapshot of an adaptive grid with four level-$1$ cells $A$,$B$,$C$,$D$. Constructing a differentially private AG requires two steps. First, the noisy counts $N'$ of $A$,$B$,$C$,$D$ are computed by adding calibrated random Laplace noise~\cite{dwork2006differential}. Second, based on the noisy counts, level-$1$ cells are further split into level-$2$ cells. Cell $D$, which has a higher noisy count of $200$ is partitioned according to a $3 \times 3$ grid, while the granularity for other cells is $2 \times 2$. Thereafter, AG adds to each level-2 cell ($c_i$, $i=1\ldots 21$) calibrated random Laplace noise. Finally, their corresponding noisy counts $n_{c_i}$ are published together with the structure of the AG.


Although AG yields small errors for general spatial queries, it is not directly applicable to SC due to its rigidity in choosing parameters. Specifically, the granularity $m_2$ of the level-$2$ grid is too coarse, leading to large geocast areas and high communication overhead. Thus, the AG method is extended to address the specific requirements of the SC framework. Particularly, a heuristic is proposed to increase the granularity $m_2$ in order to decrease overhead, but only to the point where there is at least one worker in a cell~\cite{To2014}.

\paragraph{\textbf{Task Assignment on Sanitized Data}}

On top of the noisy data, to ensure that task assignment has a high success rate, analytical models that consider task completion rate, worker travel distance and system overhead are developed. When a request for a task $t$ is posted, the SC-server queries the PSD and determines a geocast region $\mathit{GR}$ where the task is disseminated. The goal is to obtain a high success rate for task assignment, while at the same time reducing the worker travel distance $\mathit{WTD}$ and request dissemination overhead $\mathit{ANW}$.

\subparagraph{Acceptance Rate and Analytical Utility Model}

Travel distance is critical in SC, as workers need to physically visit the task locations.
A worker is more willing to accept nearby tasks~\cite{Kazemi2012}, so acceptance rate is modeled as a decreasing function of travel distance.
Also, we denote by \textit{acceptance rate (AR)} the probability $p^a (1\le p^a \le 1)$ that a worker agrees to complete a task for which he has received a request.
Thereafter, an analytical {\em utility} model is developed that allows the SC-server to quantify the probability that a task request disseminated in a certain $\mathit{GR}$ is accepted by a worker. Intuitively, the utility depends on the AR and on the worker count $\bar{w}$ estimated to be enclosed within the $\mathit{GR}$. An SC-server will typically establish an {\em expected utility} threshold {\em EU} which is the targeted success rate for a task (this is a system goal, rather than an outcome).
Generally, $\mathit{EU}$ is considerably larger than an individual worker's $p^a$, so the $\mathit{GR}$ must contain multiple workers. 

We define $X$ as a random variable for the event that a worker accepts a received task: $\mathit{P(X=True)=p^a}$ and $\mathit{P(X=False)=1-p^a}$. Assuming $w$ independent workers, $\mathit{X\sim Binomial(w,p^a)}$. We define the {\em utility} of a geocast region covering $w$ workers as:
\begin{equation}
\label{eq:utility}
U=1-(1-p^a)^w
\end{equation} 
$U$ measures the probability that at least one worker accepts the task. 
The utility definition can be extended to the case of redundant task assignment, where multiple workers are required to complete a task~\cite{To2016b}.

\subparagraph{Geocast Region Construction}

The third step in the framework is the construction and dissemination of GR. By the nature of the DP protection model, fake entries may need to be created in the worker PSD. Thus the SC-server cannot directly contact workers, not even if pseudonyms are used, as establishing a network connection to an entity would allow the SC-server to learn whether an entry is real or not, and this breach privacy. To address this challenge, the geocast mechanism was introduced for the task request dissemination. Geocast is a routing and addressing method, which is used to deliver information to all devices situated within a geographical area. Once a PSD partition is identified by the analytical model outlined above, the task request is geocast to all the workers within that partition.

Particularly, given task $t$, the GR construction algorithm must balance two conflicting requirements: determine a region that {\em (i)} contains sufficient workers such that task $t$ is accepted with high probability, and {\em (ii}) the size of the geocast region must be small. The input to the algorithm is task $t$ as well as the worker PSD, consisting of the two-level AG with a noisy worker count for each grid cell.
The algorithm chooses as initial $\mathit{GR}$ the level-$2$ cell that covers the task, and determines its $U$ value. As long as utility is lower than threshold $\mathit{EU}$, it expands the $\mathit{GR}$ by adding neighboring cells. Cells are added one at a time, based on their estimated increase in $\mathit{GR}$ utility. Following the task localness property, we take into account the distance of each candidate neighboring cell to the location of $t$, and give priority to closer cells. The algorithm stops either when the utility of the obtained $\mathit{GR}$ exceeds threshold $\mathit{EU}$, or when the size of $\mathit{GR}$ is larger than a particular threshold; hence, utility can no longer be increased. The $\mathit{GR}$ construction algorithm is a greedy heuristic, as it always chooses the candidate cell that produces the highest utility increase at each step.
The experimental results show that workers' location privacy is protected without compromising performance, and the extra travel cost is tolerable---a 20\% increase when compared to the non-private case.

Next, we present various extensions of the worker PSD, followed by an approach toward PSD for moving workers.

\paragraph{\textbf{Extensions and Enhancements of Worker PSD}}

There have been recent studies that adopt the privacy model used in~\cite{To2014}, assuming a trusted CSP and differentially private location sanitization. 
Particularly, Gong et al.~\cite{Gong2015} propose a framework that can protect the workers' location privacy when allocating tasks to the workers. Similar to~\cite{To2014}, they develop analytical models and task allocation strategies that balance privacy, utility, and system overhead. In~\cite{Gong2015}, the CSP not only aggregates workers' locations but also their reputation information, which is used to provide quality control over the reports. Consequently, a new structure called reputation-based PSD is proposed to partition the space based on both reputation and location information.

Another work studies reward-based spatial crowdsourcing that enables task assignment with optimized reward allocation (Zhang et al.~\cite{Zhang2015}). The authors also reuse the privacy framework introduced in~\cite{To2014}, in which the SC-server and workers are connected by a trusted CSP. However, unlike~\cite{To2014} that uses the adaptive grid to releases a sanitized location view to the SC-server, this study constructs a contour plot to represent the spatial distribution of workers aiming to introduce less noise than the prior technique. The contour plot is used to perform task assignment. The objective of task assignment is to find the minimum radius $r$ to ensure that the $\mathit{ASR}$ of a task is equal to {\em expected utility} threshold {\em EU}, i.e., the probability that at least one worker performs the task is no less than the threshold. 

\paragraph{\textbf{Protection for Dynamic Workers' Locations}}

Previous perturbation techniques~\cite{To2014,Gong2015,Zhang2015} assume a static scenario where workers' locations do not change. However, SC systems receive continuous requests for task assignment. Hence, it is important to keep track of the whereabouts of {\em moving} workers and to release a {\em sequence} of worker PSDs that allow effective spatial task assignment over multiple timestamps. 
The challenge is that as workers move, new snapshots of sanitized worker locations must be disclosed to maintain task assignment effectiveness. However, access to sequential releases gives an adversary more powerful attack opportunities. To counter such threats, differential privacy requires more noise injection, which in the worst case may reach amounts that are proportional to the length of the released location history (i.e., the number of disclosed snapshots). Clearly, such large noise would render the data useless, since SC is likely to be a continuously offered service in practice.
A recent study~\cite{To2016b} extends~\cite{To2014} to address the challenge of moving workers by investigating privacy budget allocation techniques across consecutive releases, and employing post-processing techniques based on Kalman filters to reduce the inaccuracy introduced by addition of noise.

\subsubsection{Encryption Techniques}

In this section we discuss studies that use encryption-based approaches. In~\cite{Shin2011} the identity and location (i.e., IP address) of workers are hidden from TS through multiple Tor relays using Onion encryption. However, Tor does not try to protect against an attacker who can see or measure both traffic going into the Tor network and also traffic coming out of the Tor network---for example, the end-to-end timing correlation attack. Thus, to prevent TS from performing a timing attack by linking multiple task requests, the workers connect to TS at random intervals. Furthermore, during tasking the workers make sure that TS does not tamper the task request from RS; otherwise, the workers can report TS as fraudulent to RS.

Shin et al.~\cite{Shin2011}, however, focus on the pull mode, which likely results in suboptimal task assignment.
Therefore, a recent study~\cite{Shen2016} proposes a secure task-assignment protocol to protect worker location privacy in the push mode. The privacy framework used in~\cite{Shen2016} is similar to~\cite{To2014} (Figure~\ref{fig:framework}), except the CSP is replaced by a \emph{privacy service provider} (PSP)---a semi-honest (i.e., honest-but-curious) third party to provide privacy functionality and collect encrypted data from workers, including encrypted location reports. With the framework, the SC-server needs to perform worker-task matching in the encrypted domain. Particularly, given a task, the SC-server communicates with PSP in the encrypted domain to find the worker with minimum travel cost to the task. The travel cost is evaluated in terms of worker-task distance and the degree of interest of the worker to the task.

The advantage of the proposed protocol is twofold. The framework is not relying on a trusted-third-party and is robust to semi-honest adversaries. Also, the privacy guarantees hold for moving workers. However, when compared to the cloaking and perturbation techniques, cryptographic-based approaches may incur higher computation overhead. In addition, the semi-honest adversary model is restrictive in terms of privacy protection and may not always hold in the real-world SC applications. That is, SC-server and PSP may not follow the specified protocol, or requesters can be malicious. Thus, a stronger privacy protocol that is resilient to malicious adversary model needs to be developed.



\section{Conclusion and Future Directions}
\label{sec:conclude}
With the popularity of mobile devices, spatial crowdsourcing is rising as a framework that enables
human workers to solve tasks in the physical world. With spatial crowdsourcing, requesters outsource a set of spatiotemporal tasks to a set of workers, i.e., individuals with mobile devices that perform the tasks by physically traveling to the specified locations of interest. However, current solutions require a worker to disclose his location to the server and/or to other requesters even before accepting a task---or a requester to disclose his tasks' locations, which can be used to infer his own location, to untrustworthy entities. In this chapter we identified the privacy threats to both workers and requesters in the two main phases of crowdsourcing: task assignment and task reporting.

We surveyed some of the most notable solutions proposing various privacy techniques, ranging from pseudonym, cloaking, perturbation to exchange-based and encryption-based approaches. 
These studies have shown encouraging results in protecting the privacy of both workers or requesters in spatial crowdsourcing. However, none of these studies address the full spectrum of threats to both entities in the push mode. Therefore, protecting the privacy of workers and requesters, simultaneously, is an open problem that is challenging and may require both encryption and perturbation-based techniques for secure and efficient tasking in spatial crowdsourcing.









\bibliographystyle{abbrv}
\bibliography{referenc}

\end{document}